\documentclass[prx,twocolumn,aps,epsf,showpacs,superscriptaddress,longbibliography,nofootinbib,notitlepage]{revtex4-2}
\usepackage[pdftex]{graphicx}

\usepackage[normalem]{ulem}
\usepackage{xfrac}
\usepackage{dcolumn}
\usepackage{bm}
\usepackage{epsfig}
\usepackage{latexsym} 
\usepackage{amsmath,mathtools}
\usepackage{amssymb}
\usepackage{color}
\usepackage{array}
\usepackage{bbm}
\usepackage{color}
\usepackage{array}
\usepackage{cancel}
\usepackage{makecell}
\usepackage[dvipsnames]{xcolor}

\usepackage[colorlinks=true,allcolors=blue]{hyperref}%

\usepackage{array,multirow}
\newcolumntype{L}[1]{>{\raggedright\let\newline\\\arraybackslash\hspace{0pt}}m{#1}}
\newcolumntype{C}[1]{>{\centering\let\newline\\\arraybackslash\hspace{0pt}}m{#1}}
\newcolumntype{R}[1]{>{\raggedleft\let\newline\\\arraybackslash\hspace{0pt}}m{#1}}

\usepackage{titlesec}
\titleformat{\subsubsection}[runin]
        {\itshape}
        {\thesection.\thesubsection.\thesubsubsection)}
        {4pt}
        {}
        [~---]
\titlespacing*{\subsubsection}{12pt}{10pt}{4pt}

\usepackage[normalem]{ulem} 

\newcommand{\<}{\langle}

\renewcommand{\>}{\rangle}
\renewcommand{\(}{\left(}
\renewcommand{\)}{\right)}
\renewcommand{\[}{\left[}
\renewcommand{\]}{\right]}
\renewcommand{\v}[1]{\boldsymbol{#1}} 

\newcommand{\tr}{\text{tr}}

\newcommand{\mie}{{\rm MIE}}
\newcommand{\mii}{{\rm MII}}

\usepackage{enumitem}
\usepackage[mathscr]{euscript}

\usepackage{tikz,tikz-cd,contour}
\newcommand{\flux}[2]{\begin{tikzpicture}[baseline=(current bounding box.center)]
	\draw (0, 0) --node{#2} (0, #1); 
	\draw (0, #1) --node{#2} (#1, #1); 
	\draw (#1, #1) --node{#2} (#1, 0); 
	\draw (#1, 0) --node{#2} (0, 0); 
\end{tikzpicture}}

\newcommand{\starterm}[3]{\begin{tikzpicture}[baseline=(current bounding box.center)]
	\draw(0, -#1) --node[pos=0.2]{#3} node[pos=0.8]{#3}(0, #1)  ; 
	\draw(-#1, 0) --node{#2} node[pos=0.2]{#3} node[pos=0.8]{#3} (#1, 0); 
\end{tikzpicture}}
\makeatletter
\def\l@subsubsection#1#2{}
\makeatother
\renewcommand{\eqref}[1]{Eq.~(\ref{#1})}

\begin{document}
\title{Universal structure of measurement-induced information in many-body ground states}

\author{Zihan Cheng}
\affiliation{Department of Physics, University of Texas at Austin, Austin, TX 78712, USA}

\author{Rui Wen}
\affiliation{Department of Physics and Astronomy, and Stewart Blusson Quantum Matter Institute, University of British Columbia, Vancouver, BC, Canada V6T 1Z1}

\author{Sarang Gopalakrishnan}
\affiliation{Department of Electrical and Computer Engineering,
Princeton University, Princeton, NJ 08544, USA}

\author{Romain Vasseur}
\affiliation{Department of Physics, University of Massachusetts, Amherst, MA 01003, USA}

\author{Andrew C. Potter}
\affiliation{Department of Physics and Astronomy, and Stewart Blusson Quantum Matter Institute, University of British Columbia, Vancouver, BC, Canada V6T 1Z1}

\begin{abstract}
Unlike unitary dynamics, measurements of a subsystem can induce long-range entanglement via quantum teleportation. 
The amount of measurement-induced entanglement  or mutual information depends jointly on the measurement basis and the entanglement structure of the state (before measurement), and has operational significance for whether the state is a resource for measurement-based quantum computing, as well as for the computational complexity of simulating the state using quantum or classical computers.
In this work, we examine entropic measures of measurement-induced entanglement (MIE) and information (MII) for the ground-states of quantum many-body systems in one- and two- spatial dimensions. From numerical and analytic analysis of a variety of models encompassing critical points, quantum Hall states, string-net topological orders, and Fermi liquids, we identify universal features of the long-distance structure of MIE and MII that depend only on the underlying phase or critical universality class of the state.
We argue that, whereas in $1d$ the leading contributions to long-range MIE and MII are universal, in $2d$, the existence of a teleportation transition for finite-depth circuits implies that trivial $2d$ states can exhibit long-range MIE, and the universal features lie in sub-leading corrections. We introduce modified MIE measures that directly extract these universal contributions. 
As a corollary, we show that the leading contributions to strange-correlators, used to numerically identify topological phases, are in fact non-universal in two or more dimensions, and explain how our modified constructions enable one to isolate universal components.
We discuss the implications of these results for classical- and quantum- computational simulation of quantum materials.
\end{abstract}

\maketitle 
\section{Introduction}
Entanglement is a fundamentally quantum phenomenon, that serves as the resource for quantum computation, and offers a powerful lens through which to classify ground-state and dynamical phases and critical phenomena in quantum many-body systems.  The universal structure of entanglement in quantum many-body ground-states offers a quantum information theoretic fingerprint of various quantum orders and critical phenomena. Since local Hamiltonian or quantum circuit dynamics can generate and propagate entanglement only at a finite speed, the long distance structure of entanglement is stable under such dynamics and is therefore useful for characterizing and classifying quantum phases~\cite{hasting2005quasiadiabatic,chen2010local}.
By contrast, quantum measurement, an inherently stochastic and non-unitary operation, can instantly generate non-local entanglement -- as famously illustrated by quantum teleportation. This capability allows short-depth circuits with measurements to generate certain types of long-range entangled states~\cite{piroli2021quantum,verresen2022efficiently,tantivasadakarn2022longrange,tantivasadakarn2023hierarchy,tantivasadakarn2023shortest,zhu2023nishimori,lee2022decoding,lu2022measurement}.
Furthermore, pre-existing entanglement in many-body states can also serve as a computational resource~\cite{raussendorf2001a,gross2007novel,gross2007measurement,vandennest2007fundamentals,doherty2009identifying,miyake2010quantum}, and there is a close connection between the universal phase structure of the pre-measured state and its computational power~\cite{stephen2017computational,raussendorf2019computationally}.

In this work, we investigate the connections between the phase of a various quantum many-body ground-states in $1d$ and $2d$, and the amount of measurement-induced long-range entanglement between distant regions, quantified by two related quantities: measurement-induced entanglement~\cite{lin2023probing} (MIE) and measurement-induced information (MII), which we define below.
The paper is organized as follows. We begin by reviewing and discussing the operational significance of MIE and MII and related quantities for various classical and quantum computational tasks. 
We then conduct a numerical and analytic investigation of the universal features of MIE and MII in a variety of $1d$ and $2d$ phases of matter and critical points through a mixture of numerical and analytic methods. We focus on states that have an efficient classical description through exactly solvable models, including free-fermion descriptions and stabilizer states. Despite their computational simplicity, this class of states encompasses a wide range of long-range entangled states including quantum critical points, topologically ordered states, and Fermi liquids, and provides a rich set of examples.

In $1d$, we show that the leading long-range contribution to MIE and MII have universal scale-invariant behavior at both conformal and non-conformal (strongly-disordered) quantum critical points.
By contrast, in $2d$ (or higher-$d$), we argue that the leading contribution to MIE is generally non-universal, due to the existence of long-range measurement-induced teleportation properties of short-depth $2d$ circuits, and discuss implications for other closely-related quantities such as strange correlators~\cite{you2014wave}. Despite this, we find that the sub-leading corrections to MIE and MII do exhibit universal features of topological orders and Fermi surfaces, and define modified notions of MIE that can directly extract these universal components.

\subsection{Definitions}
Measurement induced entanglement (MIE) is a tripartite measure of how measurements affect the entanglement structure of a state, $|\psi\>$. Partition space into three regions: $A$, $B$, and their complement $M=(AB)^c$. Consider measuring $M$ in a fixed basis $b$ with outcome $m$, which results in the post-measurement state $|\psi_m\>$ which occurs with Born probability $p_m$. Then, we define the measurement induced entanglement, $\mie_b(A, B)$ of region $A$ to be the entanglement entropy $S(A)[|\psi_m\>]$ of the post-measured state averaged over measurements:
\begin{align}
\mie_b(A, B)[|\psi\>] = \sum_m p_m S(A)[|\psi_m\>].
\end{align}
Here we have included a subscript $b$ to emphasize that the result depends (often qualitatively) on the choice measurement basis. We will occasionally drop the measurement-basis subscript where it is clear from context.

Despite its title, MIE, does not necessarily reflect entanglement that is ``induced" by the measurement, but also may capture preexisting entanglement in the initial state before measurement. 
To isolate the effects of measurement, we also define the measurement induced information (MII):
\begin{align}
\mii_b(A,B) = \sum_m p_m I(A,B)[|\psi_m\>] - I(A,B)[|\psi\>]
\end{align}
where mutual information, $I$, is defined as:
\begin{align}
I(A,B) = S(A)+S(B)-S(AB).
\end{align}
Unless otherwise specified, we will consider the case where $ABM$ is in a pure state, in which case the average mutual information after measurement is simply related to the MIE as $\sum_m p_m I(A,B)[|\psi_m\>] = 2\mie(A, B)$. 

MII probes the average amount of mutual information between $A$ and $B$ that is induced by the measurement, but did not exist before the measurement. Note that this quantity can be either positive, zero, or negative depending on the state and measurement basis. 
Measurements can increase information, resulting in positive MII, for example, when bipartite information purely between $A$ and $M$ and $M$ and $B$ gets ``teleported" by the measurement into entanglement between $A$ and $B$. A simple example of this arises in a system with four site spin-1/2 chain, with $A=\{1\}$, $M=\{2,3\}$ and $B=\{4\}$,  $|\psi\>$ is a product of singlets on sites 1,2 and 3,4 respectively, and $M$ is measured in the Bell basis of sites $2,3$.
However, measurements can also reduce or collapse entanglement. A simple example is a three qubit GHZ state: $(|000\>+|111\>)/\sqrt{2}$ which has $I=2\log 2$ information between any pair of qubits before measurement, but collapses to a product state upon measuring any of the three qubits in the computational ($Z$) basis: $\mii_Z=-2\log 2$ (whereas $\mii_X = 0$ for this example).

\subsection{Significance for quantum and classical computing}
%
In quantum information processing settings, MIE and closely-related quantities characterize how pre-existing quantum correlations in the state can be used as a resource for generating entanglement by measurements. MIE and MII also have implications for the quantum and classical complexity of describing a quantum state. Here, we briefly review and describe the operational significance of measures of measurement-induced entanglement for computational tasks.

\paragraph{Localizable entanglement:} 
The localizable entanglement (LE)~\cite{verstraete2004entanglement} is defined as the maximum over measurement bases of the MIE: ${\rm LE}(A,B) = \sup_b \mie_b(A,B)$, for the special case where regions A and B are single sites. LE upper-bounds correlation functions, thereby enabling the definition of an entanglement length scale in many-body systems that can probe non-classical correlations and has an operational meaning for contexts such as building quantum repeaters for quantum networks where one wishes to concentrate entanglement of a multipartite state into two subsystems~\cite{verstraete2004entanglement}. The maximization in the definition of LE makes it difficult to compute, and it is most useful for establishing bounds.

\paragraph{Measurement based quantum computing (MBQC):} 
MIE also partially characterizes the utility of a state for MBQC~\cite{briegel2009measurement}, where (adaptive) measurements on an entangled resource state are used to propagate and process quantum information. Long-range MIE between distant regions $A,B$ of a state is clearly a necessary condition for having a measurement-propagable computational subspace in MBQC. However, long range MIE is not a sufficient condition for MBQC as it does not address whether universal computations can be performed on the propagated information via adaptively chosen sequence of measurements in region $M$. For example, Haar random states have long range MIE between any subregions $A,B$, yet are well-known to be useless for MBQC based on single qubit measurements~\cite{bremner2009random,gross2009most}.

\paragraph{Measurement-induced phase transitions:}
Measurements can also induce phase-transitions in the post-measured trajectories, $|\psi_m\>$~\cite{li2018quantum,li2019measurement,skinner2019measurement,chan2019unitary-projective,gullans2020dynamical,choi2020quantum,zabalo2020critical,noel2022measurement-induced,koh2023measurement-induced}. In particular, $2d$ random circuits were shown to exhibit a phase transition between short- and long- range MIE tuned by the circuit depth~\cite{bao2022finite} (the teleportation fidelity order parameter used in this work is precisely the same as MIE and MII). This phenomenon was dubbed a teleportation phase transition, and has since been realized experimentally in superconducting qubit quantum processors~\cite{ai2023measurement}. The existence of long-range MIE in ``trivial" states of matter (related by a finite-depth circuit to an unentangled product state) will play an important role in our discussion of universality of MIE and MII below. Analogous to the absence of symmetry-breaking order in low-dimensions, this teleportation phase transition is believed to be possible only in two or more spatial dimensions for finite-depth local circuits.

\paragraph{Sign problem for Monte Carlo sampling:} Recent work~\cite{lin2023probing} showed that a sign problem for Monte Carlo sampling amplitudes of a state in the $b$ basis arises when the MIE is larger than the pre-measurement mutual information: $\mie_b(A, B)> I(A,B)[|\psi\>]$. We note that $\mie(A,B)-I(A,B)$ differs from $\mii$ since for a pure state of AB $I(A,B)=2S(A)=2S(B)$.  This observation directly relates the MIE to the complexity of classical simulations of quantum states.

\paragraph{Strange correlators:} MIE also arise in analytic and numerical probes of topology of a state that arise in the so-called strange correlators~\cite{you2014wave}: 
\begin{align}\label{eq:strangecorrelator}
	\frac{\<\psi_m|O_AO_B|\psi\>}{\<\psi_m|\psi\>}
\end{align}
where $O_{A,B}$ are (charged) local operators in regions $A,B$ respectively, and $|\psi_m\rangle$ is a product state, for example given by the result of measuring the system in a given single-site basis. In a path integral representation, strange correlators for topological states $|\psi\>$ are related to correlators of $O_A,O_B$ at the edge of $|\psi\>$, and hence probe the presence of topological edge modes. Just as ordinary mutual-information between regions $A$ and $B$ sets an upper bound for correlations between local operators in these regions, the MIE upper bounds the average strange correlators of $|\psi_m\rangle$ (See Appendix \ref{ap:strangecorrelatorMIE}). 
Below, we will show that MIE and conventionally-defined strange-correlators are potentially dominated by \emph{non-universal} contributions in two and higher dimensions, and introduce modified definitions of these that extract the leading universal components.

\paragraph{Complexity of quantum and classical tensor network calculations:} Tensor network states (TNS) provide efficient compressed representations of low-entangled states, such as the ground states of many local Hamiltonians. TNS wave-function amplitudes are expressed as a contraction of virtual bond degrees of freedom: $\<s_1,s_2,\dots s_N|\Psi\> = C\[T^{s_1}T^{s_2}\dots T^{s_N}\]$, where $T^{s}_{ijk\dots}$ are tensors with $s$ representing the physical degree of freedom on each site,  $i,j,k,\dots=1\dots \chi$ representing the virtual bond space, and $C$ denoting summing over virtual indices.
As explained below, the MIE structure of a state relates to measurement-induced phase transitions (MIPTs) in the bond-space of certain tensor network descriptions of the state, which informs both the design principles for quantum circuit-based tensor network calculations on quantum processors, and also the classical complexity of computing properties of tensor network states.

Classical~\cite{ferris2012perfect} and quantum-circuit based~\cite{haghshenas2022variational} methods for sampling from tensor network state (TNS) wave-functions  often involve simulating the transfer-matrix ``dynamics" of the virtual (bond) space of a codimension-one subsystem. For example a standard method to contract $2d$ TNS is to represent the first row of the TNS as a $1d$ matrix product state (MPS) in bond-space, and contract the network by evolving this MPS under the action of the row transfer matrices.
For isometric TNS (isoTNS), quantum algorithms for materials simulation have also been introduced~\cite{foss-feig2021holographic} and demonstrated~\cite{chertkov2022holographic}, in which quantum circuit dynamics together incorporating mid-circuit measurements are used to simulate the non-unitary transfer matrix dynamics.
In this context, the exponential of the MIE between distant co-dimension-one slices, $A,B$ of the TNS reflects the typical classical memory to sample wave-function amplitudes $\<s_1,s_2,\dots s_N|\Psi\>$, with $s_1\dots s_N$ representing the measurement outcomes. Similarly, for isoTNS, the MIE itself represents the average (over values $s_1,\dots s_N$), quantum resources needed to sample wave-function amplitudes with a quantum computer.

Further, when for TNS with spatial dimension larger than one, the transfer matrix for calculating $\<s_1,s_2,\dots s_N|\Psi\>$ involves dynamics of a many-body system post-selected on measurement outcomes $s_{1\dots N}$, which may exhibit distinct phases with area- or volume- law entanglement (with respect to the transfer-matrix dimension), that are separated by MIPTs~\cite{napp2022efficient,bao2022finite,cheng2023efficient}. These MIPTs represent a classical computational complexity phase transition in the difficulty of contracting the TNS. We note that while such complexity phase transitions may arise for sampling wave-function amplitudes or related global properties such as strange-correlators or wave-function overlaps, for many practical purposes one is interested mainly in correlation functions of local observables that have recently been argued to not have a complexity phase transition for fixed bond-dimension~\cite{gonzalezgarcia2023random}.

\subsection{Universal structure of MIE}
These connections motivate the need to characterize the universal features of MIE and MII in quantum many-body states, particularly the ground-states of local Hamiltonians. 
For example, for short-range entangled states the MBQC power is closely connected to the underlying symmetry and topology of the phase to which the state belongs. Moreover, in quantum simulation algorithms based on classical or quantum tensor network methods, such relations could reveal how the properties (symmetry, topology, correlation length, operator scaling dimensions, etc...) of a state that one wishes to simulate inform the circuit design principles for its quantum tensor network representation, or the complexity to perform calculations with classical tensor network methods.

Specifically, we aim to understand: what features of MIE and MII for ground-states are \emph{universal}: i.e. which are insensitive to perturbations to the parent Hamiltonian that do not drive a phase transition, or equivalently, which cannot be altered by a short-depth quantum circuit, and how are these measurement-induced entanglement features related to the phase or universality class of the state in question.
To this end, we explore, through analytic and numerical methods, the universal structure of MIE and MII in a variety of states with various types of topological orders or criticality. In particular, we emphasize the difference between the universality in $1d$ and higher dimensions, where a measurement-induced entanglement phase transition can occur at a finite circuit depth, while pre-measurement states remain in the same phase (See Sec.\ref{sec:2dnonuniversal}). In the presence of such nonuniversal long-range behavior, the universal properties only show in the subleading contribution. 
The main results of the paper are summarized in the Table \ref{tab:exponents}.
\begin{table}[!h]
	\begin{center}
		\begin{tabular}{|c| c |c| c|} 
			\hline
			System &  $I(A, B)$ &Basis $b$ & ${\rm MIE}_b(A, B)$ \\ [0.5ex] 
			\hline\hline
			1$d$ SPT & short-ranged & \makecell{symmetry \\ preserving} & $\log D$\\	
			\hline
			XX model & $\eta^{0.5}$~\cite{calabrese2009entanglement} & $ \sigma_z$ &$\eta^{0.31}$ \\ 
			\hline
			Random singlet & $r^{-2}$~\cite{ruggiero2016entanglement} & Bell & $\log2$ or $r^{-0.34}$\\
			\hline
			\makecell{Random MERA \\ with large $D$} & $\tilde{\eta}^{\log D}$ & \makecell{Arbitrary\\ local} & Extensive \\
			\hline
			Chern insulator & short-ranged & \makecell{occupation \\number} & $r^{-0.9}$ \\
			\hline
			2$d$ metal & $r^{-2}$~\cite{lepori2022mutual} & \makecell{occupation \\number} & $r^{-0.23}$ \\ [1ex] 
			\hline
			Toric code & short-ranged & $\sigma_z$ or $\sigma_x$ & $\log 2$ \\ [1ex] 
			\hline
		\end{tabular}
	\caption{Mutual information and MIE behaviors in different systems. $D$ refers to the dimension of edge states for 1$d$ symmetry-protected topological (SPT) phases and the bond dimension for random multi-scale entanglement renormalization ansatz (MERA). For the XX model and random singlet phase, the configuration of MIE is shown in the inserted figure of Fig.~\ref{fig:XX_MIE}(a) and Fig.~\ref{fig:RS_MIE}(a) respectively with $\eta=x_{12}x_{34}/x_{13}x_{24}$ and $r=x_{23}$. The configuration for the random MERA is given by Fig.~\ref{fig:MERA} with $\tilde{\eta}=x_{12}x_{34}/x_{23}x_{14}$. The configuration for the Chern insulator and 2$d$ metal is given by Fig.~\ref{fig:CI_layout}(b) , and the layout for toric code is given by Fig.~\ref{fig:TOsetup}.}
	\label{tab:exponents}
	\end{center}
	
\end{table}

The paper is structured as follows. In Sec.~\ref{sec:1dMIE} we explore the universal behavior of MIE and MII in $1d$ ground-states of both gapped, short-range correlated systems and gapless, critical systems. In Sec.~\ref{sec:2dMIE} we discuss the measurement-induced entanglement phase transition occuring at a finite circuit depth and potential adjustment of the definition of MIE and MII. Moreover, we study the universal parts of MIE and MII in topological phases and metals.

\section{MIE in $1d$ systems}\label{sec:1dMIE}
We begin by exploring universal properties of MIE and MII in $1d$ ground-states of both gapped, short-range correlated systems and gapless, critical systems. We review known structure of MIE for gapped symmetry-protected topological phases~\cite{else2012symmetry,stephen2017computational}, and selected $1+1d$ conformal field theories (CFTs) were previously explored~\cite{lin2023probing}, add additional examples of conformal and non-conformal critical points, and investigate the stability to irrelevant perturbations.

\subsection{Gapped $1d$ states}
Absent fine-tuning, typical gapped $1d$ states and generic measurement bases exhibit short range $\mie(A,B)$ that decays exponentially in distance between $A$ and $B$. This follows from the MPS representation of gapped, $1d$ states: $\<s_1,\dots s_N|\Psi\> = {\rm tr} A^{s_1}A^{s_2}\dots A^{s_N}$, where $A^{s}_{i,j}$ are $\chi\times \chi$ matrices. For unique gapped ground-states of local Hamiltonians, the matrices $A$ satisfy an injectivity property~\cite{perez-garcia2007matrix}, which guarantees short-range correlations.

A notable exception arises for the symmetry-protected topological (SPT) phases with measurements taken in an appropriate symmetry-preserving basis~\cite{else2012symmetry}, which exhibit long-range $\lim_{x_{A,B} \rightarrow \infty} \mie(A,B) = \log D$ where $D$ is the dimension of the SPT's edge states, and $x_{A,B}$ is the distance between regions $A,B$. This property is directly related to the fact that $1d$ SPTs can act as good ``quantum wires" for MBQC (i.e. are capable of coherently storing a quDit in the MBQC context). By contrast, as argued in~\cite{else2012symmetry}, measurements in generic bases lead to exponentially decaying MIE. Viewed from the perspective of the MPS transfer matrix dynamics, this short-range MIE results from mixing between the projective- and ordinary/linear- (a.k.a. ``junk") symmetry blocks of of the bond-space. We note that, despite this short range MIE for general measurement bases, it has been shown that generic MBQC operations can be performed with arbitrary target fidelity by splitting a MBQC gate into many small operations~\cite{stephen2017computational}, implemented by a gradually-evolving (and adaptively-chosen) measurement bases.

\subsection{Gapless or Critical $1d$ states}\label{ssec:gapless1d}

We next turn to gapless critical states with emergent conformal invariance, i.e. which are  described in the continuum limit by a conformal field theory (CFT).
Previous numerical investigations on select critical states of Ising-like spin chains using density matrix renormalization group~\cite{lin2023probing} demonstrated that MIE of regions $A=[x_1,x_2]$, $B=[x_3,x_4]$, depend only on the cross-ratio:
\begin{align}\label{eq:crossratio}
	\eta = x_{12}x_{34}/x_{13}x_{24}, 
\end{align}
and decayed as $\eta^\alpha$ for small $\eta$ (i.e. distance between intervals much larger than interval size).

While suggestive power-law decaying behavior does not necessarily imply universality. For example, there are other closely-related setups in which \emph{non-universal} power-law behavior can arise. Namely, Ref.~\cite{berdanier2019universal}, observed power-law dependence in Loschmidt echo of a CFT with a random time-dependent noise applied to a boundary, where the power was continuously tuned by non-universal parameters of the noise. On its face, this problem appears very different, yet, it is described by a very-similar field theory construction as MIE. Specifically, a path-integral description Loschmidt echo for the stochastically-driven boundary CFT is related by a conformal mapping to the path integral setup for calculating a second Renyi entropy version a forced-MIE in which the measurement outcomes are ``forced" to be equal to the stochastic boundary drive. A potentially important difference between the average MIE and this stochastic boundary drive, is that, in the MIE, the averaging is weighted by the Born probability of the measurement outcomes, whereas in the stochastic drive setting, it is externally imposed by the drive. Does this weighting by Born probabilities restore universality? Or could it be that the observed power-law decays of MIE for CFT states is more akin to the non-universal behavior observed in the driven boundary CFT?

\begin{figure}
	\centering
	\includegraphics[width=0.5\textwidth]{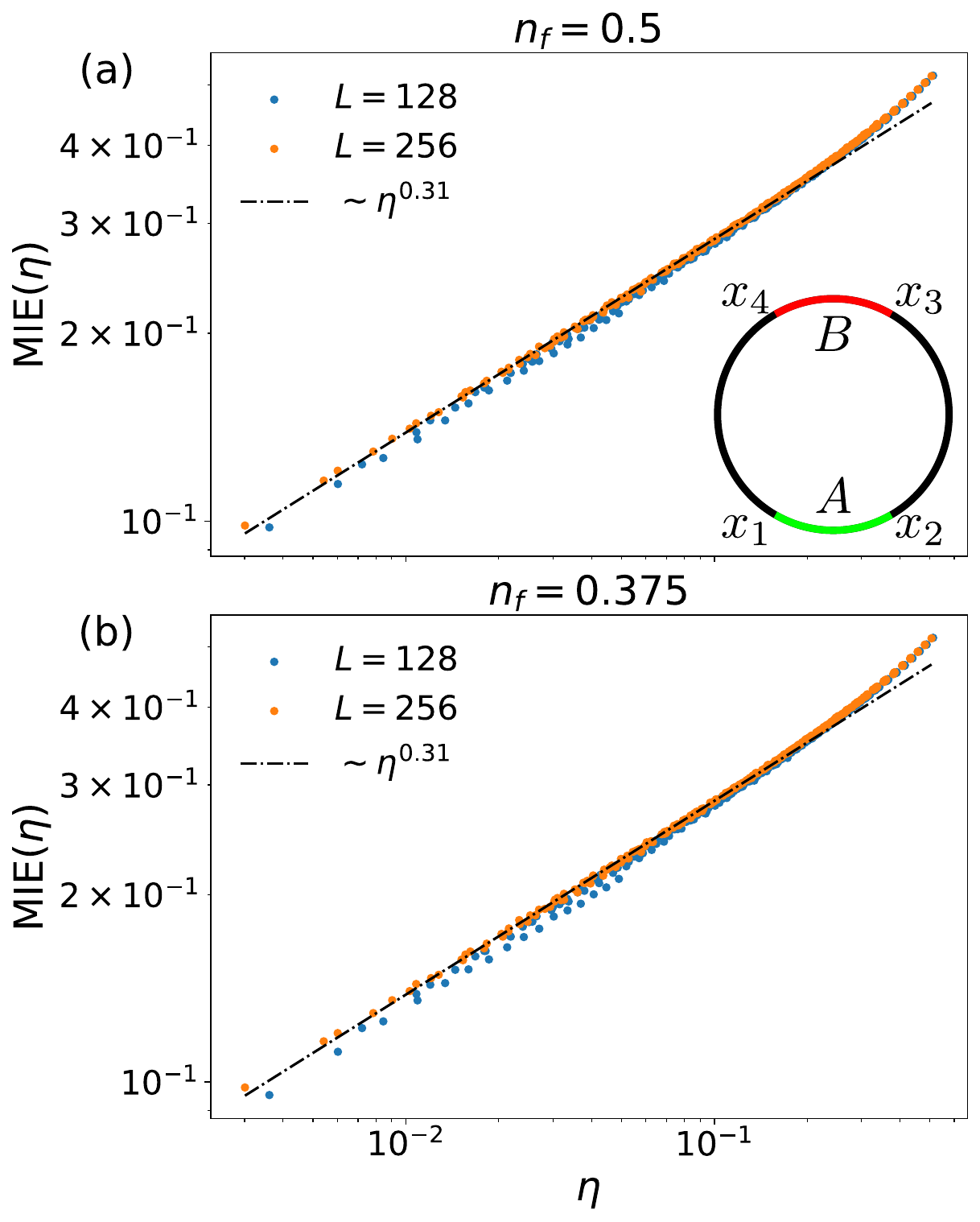}
	\caption{Data collapse of MIE in XX chain versus cross ratio $\eta=x_{12}x_{34}/x_{13}x_{24}$ with fill factor (a) $n_f=0.5$ and (b) $n_f=0.375$. Inserted figure in (a) shows the geometry of MIE regions where measured region is in black. Data are averaged over $10^6$ samples.
	}
	\label{fig:XX_MIE}
\end{figure}

\subsubsection{XX chain}
To investigate this question, we numerically study the MIE of an XX spin chain with periodic boundary condition:
\begin{align}
H_{\rm XX} = \sum_i J(\sigma^x_i\sigma^x_{i+1}+\sigma^y_{i}\sigma^{y}_{i+1})\end{align}
which can be mapped, by a Jordan-Wigner (JW) transformation, to a free fermion chain with nearest-neighbor hopping $J$ at half-filling, and exhibits Luttinger liquid ground-state described by a free-boson CFT.
Specifically we consider the ferromagnetic interaction $J=-\frac{1}{2}$, and then the Hamiltonian in free fermion language takes the form
\begin{align}\label{eq:freefermion_hamiltonian}
	H=-\sum_{i}\(c^\dagger_ic_{i+1}+c^\dagger_{i+1}c_i\) +{\rm const.}
\end{align}
with peridoic (antiperiodic) boundary condition when the total number of fermion is odd (even).
For the free fermion systems, the single orbital measurements ($\sigma_z$-basis measurements in the spin picture) can be implemented based on correlation matrix $C_{ij}=\langle c^\dagger_ic_j\rangle$ (See Appendix~\ref{ap:measurement_free_fermion}). The correlation matrix for the ground state of Hamiltonian in  ~\eqref{eq:freefermion_hamiltonian} is given by (See Appendix~\ref{ap:correlation_function} for the derivation):
\begin{align}\label{eq;correlationmatrix}
	C_{ij}=\frac{\sin\pi n_f(i-j)}{L\sin\frac{\pi(i-j)}{L}},
\end{align}
where $n_f$ is the fermion filling factor, which is $0.5$ when there is no external field. In Fig.~\ref{fig:XX_MIE} (a) we show the data collapse of MIE for $L=128$ and $L=256$ with $\eta$ over several orders of magnitude, giving $\alpha\approx0.3$, which is much smaller the $\alpha=2.$ in the forced measurement case~\cite{rajabpour2016entanglement}. Note that in the cross ratio, the finite-size effect has been taken into account by replacing $x_{ij}=\vert x_j-x_i\vert$ with the chord length $\frac{L}{\pi}\sin\(\frac{\pi \vert x_j-x_i\vert}{L}\)$.

To test the universality of power-law relation observed in MIE, we perturb the idealized XX model with chemical potential $\mu\sum_i\sigma_i^z$, which modifies the filling factor to $n_f=\frac{\arccos \mu}{\pi}$ and ~\eqref{eq;correlationmatrix} still holds. In Fig.~\ref{fig:XX_MIE} (b) the data collapse of MIE with similar power-law exponent indicates the universality of MIE. 
On the other hand, the mutual information from $1+1d$ CFT calculation scales as $\sim\eta^{\frac{1}{2}}$ at the small $\eta$ limit~\cite{calabrese2009entanglement}, thus MII is positive and dominated by the measurement-induced part at the large distance limit. 

For the measurement bases, such as $\sigma^x$, that do not conserve the fermion parity, the method in Appendix~\ref{ap:measurement_free_fermion} becomes ineffective. In Appendix~\ref{ap:DMRG}, we apply the density matrix renormalization group (DMRG) method to simulate MIE of both the XX model in the $\sigma_x$ basis and XXZ model in the $\sigma_z$ basis for smaller system sizes ($L=48$ and $L=64$). A similar exponent $\alpha\approx0.3$ is found for ${\rm MIE}_{\sigma_x}$, while for the XXZ model the exponent $\alpha$ increases along with the increasing anisotropic interaction parameter $\Delta$, which is consistent with the analytically understood increment of  
$\alpha$ in the mutual information~\cite{calabrese2009entanglement}.


\begin{figure}
	\centering
	\includegraphics[width=0.5\textwidth]{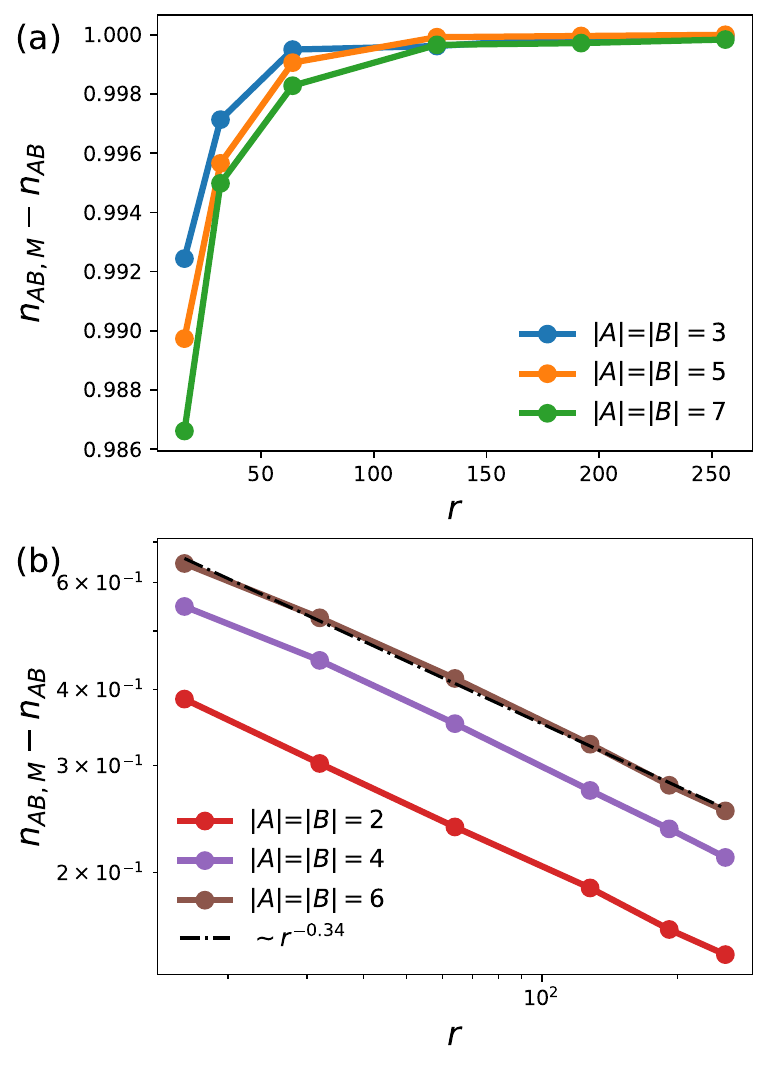}
	\caption{Number of measurement-induced Bell pairs in a random singlet state between $A$ and $B$ with $x_{23}=x_{14}=r$ for (a) odd $\vert A\vert=\vert B\vert$ and (b) even $\vert A\vert=\vert B\vert$ (geometry of the system is shown in the inserted figure of (a)). Data are averaged over $10^6$ random singlet configurations.
	}
	\label{fig:RS_MIE}
\end{figure}

\subsubsection{Random singlet phase}
Phase transitions and critical phenomena in $1d$ disordered systems often exhibit a flow to so-called infinite randomness fixed points in which the long-distance low-energy behavior are governed by rare-region effects leading to slow-glassy dynamics, strong differences between average and typical correlation functions~\cite{fisher1994random,fisher1995critical}.
A classic example arises in a $1d$ random antiferromagnetic Heisenberg spin-$\frac{1}{2}$ chain:
\begin{align}
	H = \sum_iJ_i\v{S}_i\cdot\v{S}_{i+1},
\end{align}
where $J_i\in\[0, J\]$ are random, and identically and independently distributed for each bond. Up to RG-irrelevant local dressing from quantum fluctuations, the ground-state of this model is a random singlet (RS) state in which each spin is locked into a singlet state with an another partner. The distance $r$ between a singlet pair satisfies the distribution $\sim1/r^2$. The entanglement in these long range singlets produce a log-violation of area-law with single-region entanglement is proportional to the number of singlets starting from an interval $A$ of length $l$ and terminating elsewhere~\cite{refael2004entanglement}:
\begin{align}
	S(A) = \frac{\log2}{3}\log l +k
\end{align}
where $k$ is some non-universal constant. Similarly, the mutual information between disjoint intervals $A$ and $B$ is proportional to the number of singlets spanning between $A$ and $B$~\cite{ruggiero2016entanglement}:
\begin{align}
	I(A, B) = -\frac{\log2}{3}\log(1-\eta),
\end{align}
where $\eta$ is the cross ratio defined in ~\eqref{eq:crossratio}. For the case small intervals $A$ and $B$ separated by a large distance $r$,  $I(A, B)$ scales as $\sim1/r^2$.

Now we consider the MIE and MII for such a RS state between two pre-measured disjoint intervals $A$ and $B$. Measurement in any product-state basis (e.g.  the $S^z$ basis) collapses the entanglement from all singlets between $A/B$ and $M$. The only surviving entanglement arises from singlets that directly connected $A$ and $B$ in the pre-measured state, resulting in $\mathrm{MIE}_{S^{z}}(A, B)=I(A, B)\[\vert\psi\rangle\]/2$ and $\mathrm{MII}_{S^z}(A, B)=0$.

On the other-hand, measurement in the Bell basis thus has the effect of ``teleporting'' singlets between $A/B$ and $M$ into Bell-pairs in $A \cup B$, leaving local Bell-pairs in $M$. 
We denote the number of Bell pairs connecting regions $A$ and $B$ by $n_{A,B}$. Then the MII for Bell-basis measurements of the random singlet state is proportional to the number of Bell pairs added to $A,B$ by measurement: $\mathrm{MII}_{\rm Bell} = 2\log2(n_{AB,M}-n_{A,B})$. 
 Specifically we consider the Bell measurements on nearest-neighbor spin pairs in a measurement region containing an even number of spins. For the simplest case, i.e. the un-measured regions $A$ and $B$ only contain single sites. In this case: $n_{AB, M}$ has to be $1$ since all other sites are paired after Bell measurements and no existing pair can be eliminated. In result,  $\mathrm{MIE}_{\rm Bell}=\log2$ and $\lim_{L\rightarrow\infty}\mathrm{MII}_{\rm Bell}=2\log2$. The similar long range behavior is expected for odd number of $\vert A\vert$ and $\vert B\vert$ (for random singlets, we have assumed total system size is even), which is numerically verified in Fig.~\ref{fig:RS_MIE}(a). 
However, for the cases $A$ and $B$ containing even number of site, there is no guarantee for the formation of long-range Bell pairs. 
Instead, measurements can either teleport an $A/M$ singlet into an $A/B$ singlet contributing to MII, or they might teleport the $A/M$ singlet into a local singlet with both spins in $A$, which does not contribute to MII. The ratio of these two possibilities depends in a complicated fashion on the disorder configuration and geometry of the regions, but nevertheless yields a universal scaling form. To investigate it, we perform a numerical calculation using the strong-disorder renormalization group method. We start from a random Heisenbeg Hamiltonain and apply Ma–Dasgupta rule for the renormalization of the strongest bond until obtaining a random-singlet ground state. Subsequently, we perform Bell measurements on nearest-neighboring pairs which effectively rearrange the Bell pairs. The results shown in Fig.~\ref{fig:RS_MIE}(b) demonstrate for even number of $\vert A\vert$ and $\vert B\vert$  the $\mathrm{MII}_{\rm Bell}(A, B)$ features a power-law decaying $\sim r^{-0.34}$.

\begin{figure}
	\centering
	\includegraphics[width=0.5\textwidth]{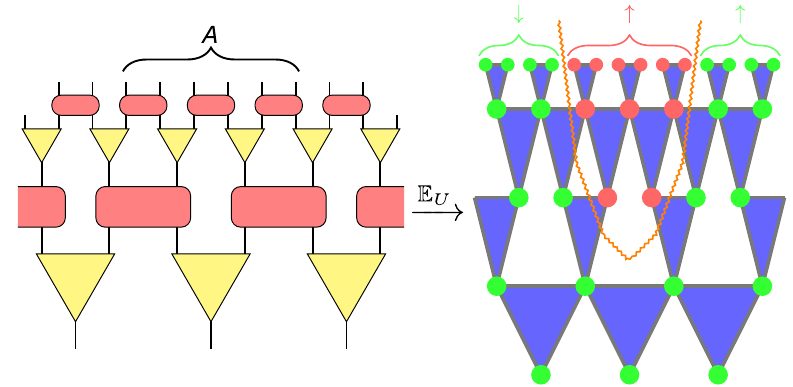}
	\caption{(Left) Schematic figure for MERA tensor networks; (Right) Mapped statistical mechanics model after Haar random average. Mini-cut domain wall is indicated by the orange wave line.
	}
	\label{fig:MERA}
\end{figure}

\subsubsection{Random MERA states}
Another tractable model for computing MIE are random multi-scale entanglement renormalization (MERA) tensor networks~\cite{vidal2008class,evenbly2009algorithms}. MERA tensor networks can produce the states with logarithmic entanglement. Consequently, they are often used to model critical ground states that are described by a CFT~\cite{pfeifer2009entanglement,milsted2018tensor}. Entanglement features of MERAs with either non-unitary Gaussian random tensors or Haar random unitaries and isometries can be analytically computed by mapping them, via a replica trick, to a classical statistical mechanics model~\cite{hayden2016holographic,vasseur2019entanglement,zhou2019emergent,bao2020theory,jian2020measurement,nahum2021measurement}. In the limit of large bond dimension, $D\rightarrow\infty$, the statistical mechanics model calculations become tractable and reduced to pure geometric quantities that are identical to the Ryu-Takanagi principle for computing entanglement via holographic field theory/gravity duality~\cite{hayden2016holographic}. 

In the stat-mech mapping~\cite{potter2022entanglement}, the entanglement entropy of a boundary region $A$ maps to the free energy in a classical statistical mechanics model of generalized ``spins'' sitting on the vertices of the tensor network, with ferromagnetic interactions of strength $J=\log D$ between spins connected by edges of the network, and boundary ``magnetic'' fields of strength $h=\log D$~\cite{hayden2016holographic,vasseur2019entanglement,jian2020measurement}, which explicitly break the replica permutation symmetry. 
For the simplicity, we describe the two-replica case where the model reduces to an Ising model with two spin configurations $\uparrow, \downarrow$ (for general number, $Q$, of replicas, the resulting model is Potts-like with $Q$ spin-flavors, but the universal features inside the ordered phase of the model at large-$D$ are expected to be independent of $Q$). 
The phase diagram of Ising models on graphs with hyperbolic geometry has been studied both analytically~\cite{wu1996ising} and numerically\cite{breuckmann2020critical}. Unlike for Euclidean geometries, where boundary conditions do not effect bulk critical properties, on hyperbolic graphs, the boundary contains an extensive fraction of the total number of sites, and the resulting phase diagram is sensitive to boundary conditions.
For fixed boundary conditions, with an explicitly symmetry-breaking field at the boundary, there is a single bulk order-to-disorder transition.
By contrast, for free-boundary conditions, the ``ordered" phase splits into two phases: a low temperature uniformly-ordered phase with a single spontaneously-chosen magnetization, and a moderate temperature phase with a finite fraction of disordered spins. 

\begin{figure}
	\centering
	\includegraphics[width=0.5\textwidth]{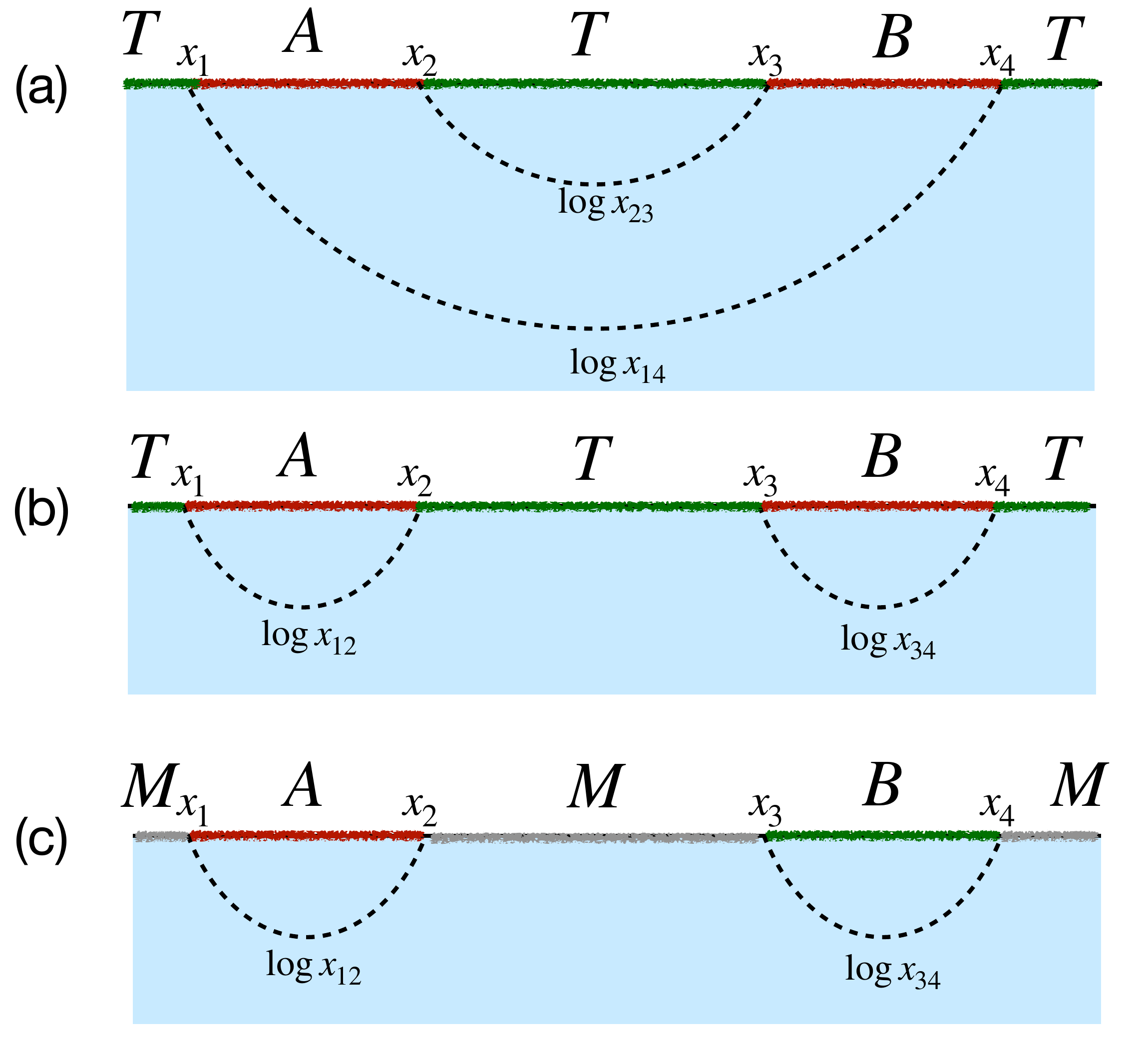}
	\caption{(a), (b) two possible minimal cuts for $S(A\cup B)$; (c) two possible minimal cuts for ${\rm MIE}(A)$.
	}
	\label{fig:adscut}
\end{figure}

In the present context, the single-region entanglement $S(A)$  is given by the free energy cost having $\uparrow$ boundary fields outside of $A$ and $\downarrow$ boundary fields in its counterpart $\bar{A}$, which forces a domain wall (DW) into the system.
Therefore, $S(A)$ corresponds to the fixed-boundary condition hyperbolic-Ising model, and exhibits a single (dis)ordering phase transition at a critical $D_c$. Below this critical bond-dimension, there is only a local cost to inserting a boundary domain wall resulting in area-law entanglement ($S(A)\sim $ constant).
Above this critical bond-dimension, the bulk is ordered, and the DW tension (energy cost per unit length) is proportional to $\log D/D_c$.
In the following we consider the limit of large $D$, where the effective spin model will be in an ordered, ferromagnetic phase.
Here, fluctuations in the DW shape are strongly penalized and the DW follows a minimal cut of the interaction edges. For a MERA tensor network with a $1d$ boundary, the minimal cut connects the ends of region $A$ interval dives down into the bulk as shown in Fig.~\ref{fig:MERA} and has length $\sim\log\vert A\vert$. which gives $S(A)\sim\log D\log\vert A\vert$.
The mutual information for two disjoint intervals $A=\[x_1, x_2\]$ and $B=\[x_3, x_4\]$ is given by the free energy of two competing cut configurations for $S(A\cup B)$
\begin{align}
	I\(A, B\) = F_A+F_B+\log\(e^{-F_A-F_B}+e^{-F_{AB}}\),
\end{align}
where $F_{A}+F_{B}\sim\log D\log x_{12}x_{34}$ (Fig.~\ref{fig:adscut}(a)) and $F_{AB}\sim\log D\log x_{14}x_{23}$ (Fig.~\ref{fig:adscut}(b)).  In the limit of $D\rightarrow\infty$, $I\(A, B\)$ shows an abrupt jump as a function of $\tilde{\eta}=x_{12}x_{34}/x_{14}x_{23}$, which relates to the cross ratio defined in ~\eqref{eq:crossratio} as $1/\eta=1/\tilde{\eta}+1$,
\begin{align}
	I\(A, B\) \sim \begin{cases}\log D \log \tilde{\eta} & \tilde{\eta}>1 \\ \tilde{\eta}^{\log D} & \tilde{\eta}<1\end{cases}.
\end{align}

We now turn to average MIE and MII for this random MERA. We note that, the random nature of the tensors means that the average MIE does not depend on measurement basis. The principal difference in the stat-mech mapping is that the measured region $M$ has free-boundary condition, and thus exhibits a \emph{distinct, separate phase transition} from that of the (fixed-boundary condition) stat-mech model for entanglement (without measurement). As a result of the free-boundary conditions, the DW ends are no longer linearly-confined to the ends of the boundary of entanglement region, but can fluctuate into measured regions. 
The cost to change the size of the minority spin domain encapsulating $A$ from the minimum domain, to $x>|A|$, scales as $\log D \log x/|A|$. This competes with an entropic gain $\sim \log x$ from the fluctuations. The competition of energetic vs. entropic logarithmic factors is familiar from Kosterlitz-Thouless (KT) transitions arising in $1d$ XY models and discrete $1d$ spin chains with long range interactions decaying as $\sim 1/r^2$ with distance $r$. By analogy, there should be a critical bond-dimension $D_c'>D_c$ at which the stat-mech model for MIE changes from an ordered phase $(D_c<D<D_c')$ with short-range MIE, to an ordered phase $(D>D_c')$ with $\mathrm{MIE}(A, B)\sim \log |A|$ as discussed above.
This phase transition is analogous to the ``finite-time" teleportation transition in $1d$ quantum circuits with power-law range gates discussed in~\cite{bao2022finite}. 
We will discuss the implications to MIE for a related teleportation transition in finite-depth $2d$ circuits below. There we will argue that the leading contribution to MIE in the teleporting phase is long-range and non-universal.

In the large-$D$ limit, such fluctuations are strongly suppressed, and thus the leading order of MIE is given by
\begin{align}
	\mathrm{MIE}(A, B) \sim\log D\log\(\min\{x_{12}, x_{34}\}\),
\end{align}
which is equal to the minimum pre-measurement of entanglement entropy in $A$ and $B$ (min-cuts are given in Fig.~\ref{fig:adscut}(c)). We note that this result holds even when the distance between $A$ and $B$ becomes very large such that the pre-measurement $I(A, B)\approx0$. In such case one has \begin{align}
	\mathrm{MII}(A, B)\approx2\mathrm{MIE}(A, B)\sim2\log D\log\(\min\{x_{12}, x_{34}\}\).
\end{align}

This behavior of MIE qualitatively differs from that found in the free fermion (XX-chain) studied numerically in the previous section. This deviation is a direct consequence of the different phase diagrams for ferromagnetic spin models on hyperbolic geometries with fixed- or free- boundary conditions. For the CFT, MIE followed the behavior of a four-point function, depending only on the cross-ratio: $x_{12}x_{34}/x_{13}x_{24}$. This suggests a possible difference in the structure of MIE between holographic tensor network states and (minimal-model) CFTs. This behavior contrasts that for the pre-measurement mutual-information, $I(A,B)$, which behaved like a four-point function for both models.

\section{MIE in $2d$}\label{sec:2dMIE}
We next turn to investigating the structure of measurement-induced entanglement in $2d$ states. Here, the existence of teleportation phase transitions in finite-depth $2d$ circuits dramatically alters the structure and universality of MIE~\cite{napp2022efficient,bao2022finite}. Nevertheless, we find that universal signatures of topology and non-local entanglement associated with Fermi surfaces still arises as subleading corrections to MIE, and discuss how to directly extract these universal signatures.

\begin{figure}
	\centering
	\includegraphics[width=0.5\textwidth]{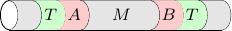}
	\caption{Schematic figure for the traced-MII: mutual information between $A$ and $B$ with green regions $T$ traced out and grey regions $M$ measured out, 
	}
	\label{fig:traced_MII}
\end{figure}

\subsection{Trivial Gapped States}\label{sec:2dnonuniversal}
A trivial gapped state is one that can be produced from a product state by a finite depth local circuit.
Unlike $1d$ systems, measurements in $2d$ (or higher dimensions) can induce nontrivial entanglement phase transition with finite-depth random circuits~\cite{napp2022efficient,bao2022finite}, i.e. after applying $t>t_c$ layers of random circuits, arbitrary initial state can feature extensive measurement-induced entanglement $\mathrm{MIE}\(A, B\)= O\(\vert A\vert\)$. Therefore, the leading order of $\mathrm{MIE}$ in $2d$ is some non-universally constant and on the other hand, universal behaviors, such power-law decaying will be overshadowed.

\subsubsection{Leading contributions to strange correlators are non-universal}
The existence of extensive term in MIE also implies a non-universal constant part in strange correlators in $2d$, that there exists two trivial states $\vert\psi\rangle$ and $\vert \psi_m\rangle$ giving long-ranged strange correlator in ~\eqref{eq:strangecorrelator} and thus overshadows the diagnosis of nontrivial topological states, which contributes a power-law decaying~\cite{you2014wave}. To show this explicitly, we consider a $2d$ trivial initial state $\vert\psi\rangle$ with application of  $t>t_c$ layers of random circuits which doesn't have pre-measurement long-range correlation but is in the ``teleportation phase''~\cite{bao2022finite}: $\mathrm{MIE}(A, B)=O\(\vert A\vert\)$. Specifically, we consider local Hilbert dimension $D=2$, $\vert A\vert=\vert B\vert = 1$ separated by distance $r$ and the rest degrees of freedom are measured out. Then the post-measurement state can be written as
\begin{align}
	\vert\psi\rangle_{AB}=\lambda_0\vert a_0\rangle\vert b_0\rangle + \lambda_1\vert a_1\rangle\vert b_1\rangle,
\end{align}
where $\vert a_{0/1}\rangle$ and $\vert b_{0/1}\rangle$ are orthonormal local basis supported by $A$ and $B$. In the ``teleportation phase'' $\mathrm{MIE}(A, B)=-\(\lambda_0^2\log\lambda_0^2+\lambda_1^2\log\lambda_1^2\) = O(1)$, which indicates $\lambda_{0/1}$ are both finite non-zero numbers. Then by choosing $O_A=\vert a_0\rangle\langle a_1\vert$, $O_B=\vert b_0\rangle\langle b_1\vert$ and $\vert\psi_m\rangle_{AB}=\vert a_0\rangle\vert b_0\rangle$, one can verify that the (connected) strange correlator between $\vert\psi\rangle$ and $\vert\psi_m\rangle=\vert\psi_m\rangle_{AB}\vert\psi_m\rangle_{M}$
\begin{align}
	\frac{\<\psi_m|O_AO_B|\psi\>}{\<\psi_m|\psi\>}-\frac{\<\psi_m|O_A|\psi\>}{\<\psi_m|\psi\>}\frac{\<\psi_m|O_B|\psi\>}{\<\psi_m|\psi\>} = \frac{\lambda_1}{\lambda_0},
\end{align}
which is generally a non-vanishing number for finite $\mathrm{MIE}(A, B)$ independent of the distance $r$. Note that the previous choice of $O_A$, $O_B$ and $\vert\psi_m\rangle_{AB}$ can be in arbitrary local orthonormal basis if we consider the typical case under random circuits, where each projector $\vert a_0/b_0\rangle\langle a_1/b_1\vert$ will typically contribute with a prefactor $O\(\frac{1}{4}\)$, and then the (connected) strange correlator will still have a finite long-range behavior.

\subsection{Isolating the universal features of MIE and strange correlators}
In this section we introduce two modified versions of MIE (which can be readily adapted to strange correlators), that isolate the universal contributions of MIE that depend only on the phase of matter for the pre-measured state, from the non-universal ones arising from the teleportation transition.

\paragraph{Scaled MIE}
To probe the universal behavior of $\mathrm{MIE}$ or strange correlator, one has to go to the subleading term with power-law decaying. Generally, the measurement-induced non-universal effect is proportional to the size of post-measurement region $A$, thus to get rid of the non-universal extensive term of MIE, one can redefine MIE as
\begin{align}
	\widetilde{\mathrm{MIE}}(A, B)=2\mathrm{MIE}_{1/2}(A/2, B/2)-\mathrm{MIE}(A, B),
\end{align}
where $\mathrm{MIE}_{1/2}(A/2)$ represents the MIE for a system scaled by $1/2$ in the particular direction of interest, not only for the region $A$. Within this definition, the extensive terms in MIE are canceled, while the sub-leading power-law relation is retained. 
A non-ideal aspect of this difference between scaled MIE is that it assumes a specific form of the subleading universal corrections. In the next section, we introduce an alternative means to isolate the universal aspects of MIE from the non-universal teleportation transition ones, that is agnostic to the precise scaling structure of the subleading universal terms.

\paragraph{Partially-traced MII}
Another possible solution is to consider the quantity that is not affected by the measurement-induced phase transition but still features the universal behaviors. One candidate, shown in Fig.~\ref{fig:traced_MII}, is the partialy-traced MII, where instead of considering the total mutual information between two post-measured regions $A$ and $B$, we consider the mutual information between two subregions $A_0$, $B_0$ of them. In this case, the left region $A/A_0\cup B/B_0$ is traced over (see Fig.~\ref{fig:traced_MII}) and thus imposes a definite spin configuration in the statistical mechanics picture which makes this quantity always vanishing in the long-range limit for trivial initial states. 

For the models considered in the remainder of the paper, we find that the ground-states naturally lie in the non-teleporting phase. For this reason, we do not need to explicitly modify MIE to subtract a non-universal extensive piece.

\subsection{Chern insulator}
\begin{figure}
	\centering
	\includegraphics[width=0.5\textwidth]{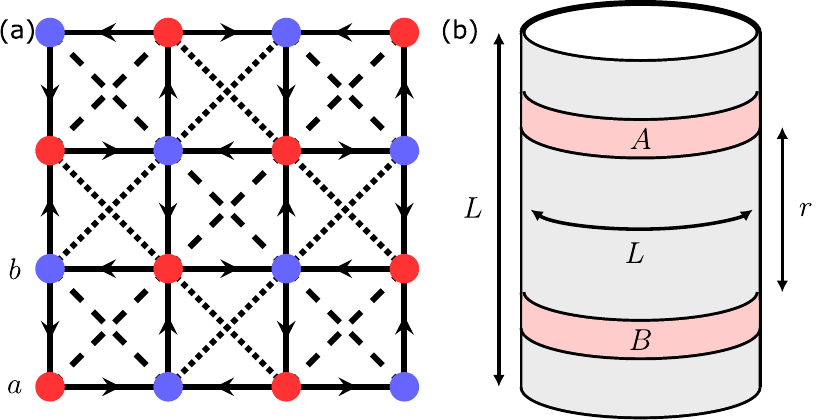}
	\caption{(a) Two band model on a square lattice, where nearest-neighbor hopping amplitudes are $t_1e^{i\pi/4}$ along the arrow direction and next-nearest-neighbor hopping amplitudes are $t_2$ and $-t_2$ along the dashed and dotted diagonals. (b) MIE on a cylinder geometry where grey region is measured out. 
	}
	\label{fig:CI_layout}
\end{figure}
\begin{figure}
	\centering
	\includegraphics[width=0.5\textwidth]{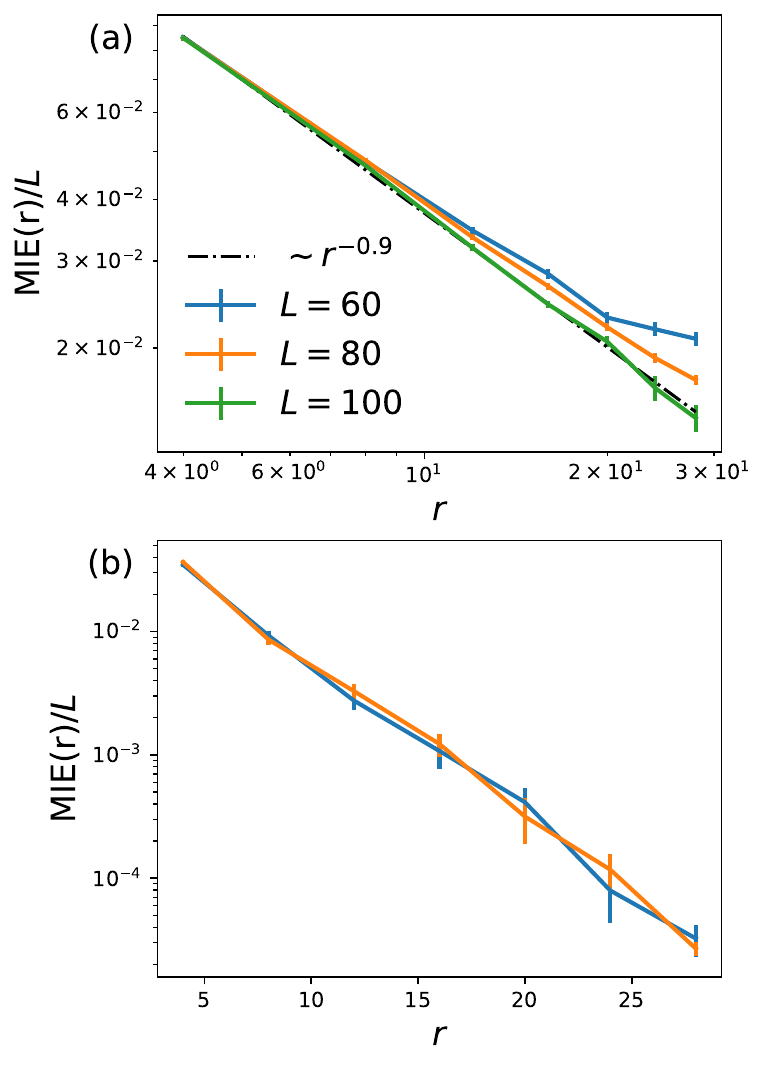}
	\caption{MIE in the two-band model with cylinder geometry for (a) $V=0$ and (b) $V=0.8$, where the result is averaged over $120$ measurement realizations.
	}
	\label{fig:CI_MIE}
\end{figure}

In Chern insulators, the strange correlator decays as a power law (in particular, for the free-fermion case $\sim r^{-1}$)~\cite{you2014wave,wu2015quantum,lepori2023strange}, which can be understood by mapping to the standard correlator in the one-dimensional CFT. Since in the CFT, MIE is lower-bounded by the averaged square of strange correlators (see Appendix.~\ref{ap:strangecorrelatorMIE}), we expect the universal part of MIE in Chern insulators also features a power-law behavior. 

To investigate this question, we numerically study the MIE in a two band model on a square lattice introduced by Ref.~\cite{neupert2011fractional}, which holds non-trivial Chern number. As shown in Fig.~\ref{fig:CI_layout}(a), the model consists of two sublattices $a$ and $b$, where nearest-neighbor hopping amplitudes are $t_1e^{i\pi/4}$ along the arrow direction and next-nearest-neighbor hopping amplitudes are $t_2$ and $-t_2$ along the dashed and dotted diagonals. To compare the results with trivial insulator, we add a staggered one-site potential, which is $V$ for $a$ sublattice and $-V$ for $b$ sublattice. The total tight-binding Hamiltonian reads
\begin{align}
	H &= -t_1\sum_{\langle i, j\rangle}e^{i\phi_{ij}}c^\dagger_{a, i}c_{b, j} + {\rm h.c.}-t_2\sum_{\langle\langle i, j\rangle\rangle}\tau_{ij}c^\dagger_{a, i}c_{a, j}\nonumber\\
	&-t_2\sum_{\langle\langle i, j\rangle\rangle}\tau_{ij}c^\dagger_{b, i}c_{b, j}+V\sum_i\(c^\dagger_{a, i}c_{a, i}-c^\dagger_{b, i}c_{b, i}\),
\end{align}
where $\langle i, j\rangle$ represents nearest neighbors and $\langle\langle i, j\rangle\rangle$ represents next-nearest neighbors.  $\phi_{ij} = \pi/4$ ($-\pi/4$) if the hopping is along (against) the direction of the arrow, and $\tau_{ij} = 1$ ($-1$) if the hopping is along dashed (dotted) diagonals. For $-4t_2/t_1<V<4t_2/t_1$, the band features Chern number $C=1$, while for $V>4t_2/t_1$ or $V<-4t_2/t_1$, the band is trivial. 

Specifically, we let $t_1=1.$ and $t_2 = 0.1$ and consider the $\mathrm{MIE}$ with measurements taken in the basis of occupation number on each site between two rings with width $2$ that are separated by distance $r$ on an $L\times L$ lattice with cylinder geometry (Fig.~\ref{fig:CI_layout}(b)). 
Since this is a free fermion system, similar numerical method discussed in Sec. \ref{ssec:gapless1d} and Appendix \ref{ap:measurement_free_fermion} can be applied.
As shown in Fig.~\ref{fig:CI_MIE}(a), the MIE for nontrivial Chern number ($V=0$) features a power-law decay $\sim r^{-0.9}$ satisfying the lower bound given by the strange correlator. As a comparison, in a trivial insulator ($V=0.8$) the MIE decays exponentially (Fig.~\ref{fig:CI_MIE}(b)). Since the ground state of Chern insulators has short-ranged pre-measurement entanglement, the MII also features power-law decaying.

\begin{figure}
	\centering
	\includegraphics[width=0.5\textwidth]{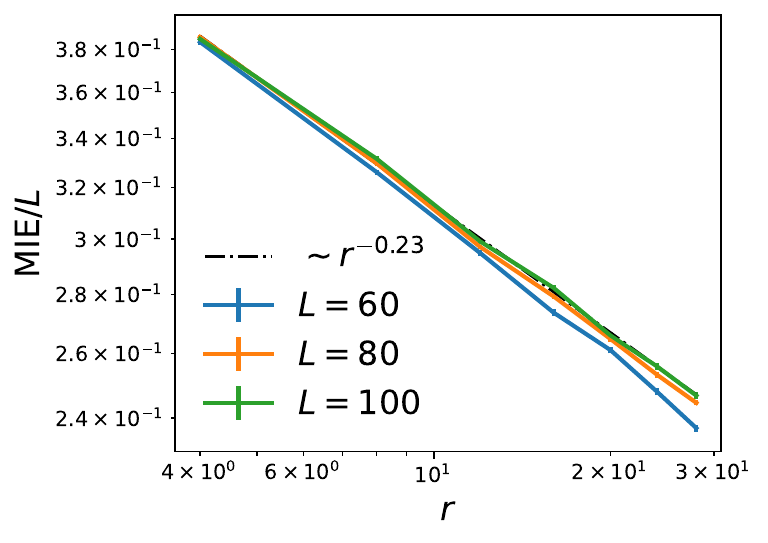}
	\caption{MIE in the $2d$ free fermion model on the square lattice with cylinder geometry, where the result is averaged over $120$ measurement realizations.
	}
	\label{fig:FF_MIE}
\end{figure}

\subsection{Metals}
As a final numerical example, we consider the MIE of a non-interacting metal, with measurements taken in the basis of occupation number on each site. Due to the presence of a Fermi-surface of gapless excitations, the metallic state has a large entanglement before measurement, with entanglement of a region of size $L$ scaling as $S\sim k_F L \log L$ where $k_F$ is the Fermi wave-vector.
We consider MIE in a free fermion tight-binding model 
\begin{align}\label{eq:metal}
	H=-\sum_{\langle i , j\rangle}\(c^\dagger_ic_{j}+c^\dagger_{j}c_i\),
\end{align}
where $\langle i, j\rangle$ represents nearest neighbors on a two-dimensional square lattice. The ground state of this Hamiltonian can be considered as a metal.

Similar to the Chern insulator case, we consider the MIE between two rings with width $2$ separated by distance $r$ on an $L\times L$ lattice with cylinder geometry (Fig.~\ref{fig:CI_layout}(b) and apply the free-fermion method discussed in Sec. \ref{ssec:gapless1d} and Appendix \ref{ap:measurement_free_fermion}. As shown in Fig.~\ref{fig:FF_MIE}, the MIE features a power-law decaying $\sim r^{-0.23}$, where corresponds to $\alpha\approx0.115$ (considering $\eta\sim r^{-2}$ in the large distance limit), which smaller than $\alpha\approx0.3$ of the $1d$ case. 
Since the pre-measurement $2d$ free fermion systems can be decoupled into $1d$ chains, the mutual information scales as $\sim\eta$~\cite{lepori2022mutual}~\footnote{Note that although JW transformation can map the XX model to $1d$ free fermion chain, they actually have different mutual information behavior due to the implicit non-local
structure from JW transformation~\cite{igloi2010on}).}. 
By contrast, the random measurement outcomes break the translation invariance, such that the MIE of the metal does \emph{not} reduce to decoupled copies of $1d$ free-fermion systems for each momentum parallel to the measurement-region boundary. In particular, the MIE of the $2d$ metal decays with a different power of $r$ than for the $1d$ free fermion system.
Since this pre-measurement mutual information decays much more rapidly than the MIE, the MII in large distance limit is dominated by the measurement-induced part and scales as $r^{-0.23}$.

\subsection{Topological orders (String-net liquids)}
While leading order of MIE in a trivial gapped state shows no universal behavior, MIE of a topologically ordered state contains a long-range (constant) term that depends only on the topological information of the measurement scheme, namely it depends only on the homological class of the measurement region and is invariant upon deformation of the measurement region. 

Measurements of a sub-region of a topologically-ordered state can be interpreted as imposing boundary conditions on the unmeasured region. To see this, observe that after performing a local measurement in a region $M$, the state in  $M$ is in a tensor product state that has no topological order while the unmeasured region remains topologically ordered. Therefore the interface between the two regions is an interface between topological order and trivial state. Boundaries of $2d$ topological orders are well-understood--they are in 1-1 correspondence with different ways of condensing anyons of the topological order. This classification and characterization of boundary conditions allow us to analyze the post-measurement state of a topologically order state systematically. 

\subsubsection{Toric code on a torus}
\begin{figure}
	\centering
	\includegraphics[width=0.5\textwidth]{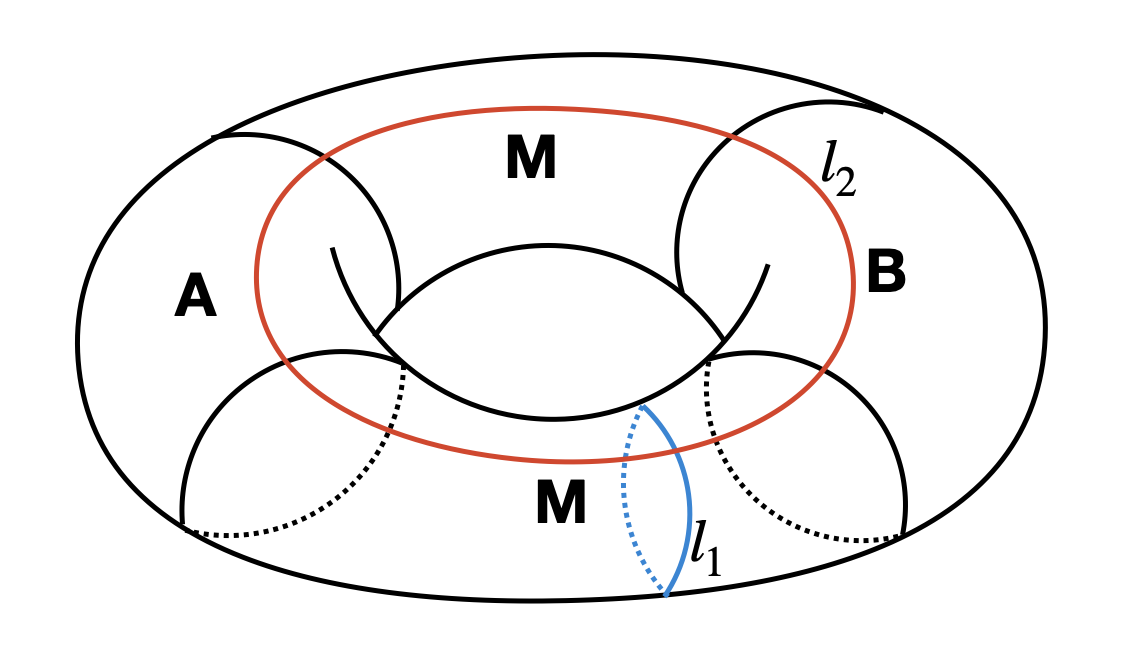}
	\caption{\textbf{MIE of toric code on a torus.} The torus is divided into regions $A,B,M$. The $M$ region is then measured out and we are interested in the MIE between $A$ and $B$. Two non-contractible loops $l_1,l_2$ are shown in blue and red.
	}
	\label{fig:TOsetup}
\end{figure}

As a warmup example let us consider the toric code, $D(\mathbb{Z}_2)$, on a torus, $T^2$, modeled is a spin $\frac{1}{2}$ degree of freedom at each link of a square lattice, with Hamiltonian:
\begin{align}
	H=&-\sum_{\starterm{0.2}{}{}}\starterm{0.7}{}{$X$}-\sum_{\flux{0.3}{}} \flux{1}{$Z$}\nonumber\\
	=&-\sum_v A_v-\sum_f B_p,
\end{align}
where $v$ and $p$ represents vertice and plaquettes, and  $Z,X$ are Pauli operators. 
If the measurement region is contractable, then due to the topological nature of the system, there is no long-range measurement-induced entanglement.
Instead we consider measuring a non-contractable region $M$, that separates two unmeasured regions $A,B$ as shown in Fig.~\ref{fig:TOsetup}. 
The boundaries between measured and unmeasured region are on non-contractible loops which we call $l_1$. The other non-contractible loop intersects $l_1$ once and is called $l_2$. 
The ground space of the toric code Hamiltonian on the torus is 4-fold degenerated and is invariant under the action of string (loop) operators on $l_1,l_2$. The ground state is invariant under application of string operators that are labelled by anyon types, $\{e,m,f=e\times m\}$. A generating set of string operators can be chosen as $e$-strings and $m$-strings. An $e$-string is a product of $Z$ operators along a path of the lattice while an $m$-string is a product of $X$ operators along a path of the dual lattice. We denote loop operators of type $a=e,m$ on loops $l_{i=1,2}$ as $W^{a}_i$. Then these operators preserve the ground space and their eigenvalues can be used to label states in the ground space. In the language of string-net condensation the ground states are equal-weight superpositions of closed $e$-loop configurations, where we define links with $X=-1$ as having an $e$-loop segment. Then the 4-fold degenerated states can be characterized by the parity of number of loops along non-contractible directions $l_1,l_2$. This way of labelling provides us with a basis of the ground space which we denote as $|\alpha,\beta\>_{SN}$ and call string-net basis, $\alpha,\beta=0,1$ are the parities of number of loops along $l_1,l_2$ respectively. 

Now let us consider starting with a state with even parity of loops long $l_1,l_2$: $|0,0\>_{SN}$ and measuing in $X$-basis in region $M$. The measurement will then project onto some fixed loop configuration in $M$, while the unmeasured region remains a equal weight superposition of various loop configurations. From the measurement outcomes we can infer the parity of loops along $l_1$ in $M$. If the measurements yield a result such that the number of loops along $l_1$ in $M$ is even, then we know that the total number of loops along $l_1$ in $A,B$ must be even since the initial state has even parity of loops along $l_1$. This condition entangles the string-net configurations in $A,B$: they can have either both even or both odd parity along $l_1$. Apart from this condition the states in $A,B$ are still equal weight superpositions of all possible loop configurations. The post-measurement state in region $A,B$ is thus a Bell pair with entanglement entropy $\log 2$. On the other hand if the measurement projects onto a state in region $M$ with odd number of loops along $l_1$, then the postmeasurement state must have opposite parities of loops along $l_1$ in $A$ and $B$. In this case the post-measurement state is again a Bell-pair and has entanglement entropy $\log 2$. Therefore after averaging over measurement outcomes with Born probability we have $\mie(A,B)=\log 2$. One can verify that starting with any of the 4 states $|\alpha,\beta\>_{SN}$ will yield $\mie(A,B)=\log 2$ if one measures in $X$-basis. 

A scheme dual to the above measurement scheme is to start with a state with definite parities of dual loops along $l_1,l_2$\footnote{One says there is a a dual string at a link if $Z=-1$ at that link, and a dual string lives on a path of the dual lattice.} and make measurements in $Z$-basis, which will also give $\mie(A,B)=\log 2$.

The above example illustrates that the $\mie$ of a topological order is a constant that is invariant under deformations of the measurement region for certain measurement basis. But several question remains. How exactly is the $\mie$ related to the data of the topological order such as quantum dimension of anyons? Can the result for toric code be generalized to other topological orders? We address these questions by reformulating the above calculation in a way that makes it suitable for generalizaiton, along the way these questions will be resolved automatically. 

\subsubsection{Minimally entangled states}
 Apart from the string-net basis $|\alpha,\beta\>_{SN}$ discussed before, there are other basis for the ground space of a topological order on a torus.  An important basis consists of the minimally entangled states (MES)~\cite{Zhang_2012,Wen_2016}. MES will be crucial in obatining the MIE for a general ground state of abelian quantum double models. Here we briefly summarize their construction. Consider all loop operators on a non-contractible loop $l_1$, these are generated by $W^e_1,W^m_1$ as an algebra. Since these two operators commute, we can diagonalize them simultaneously within the ground space. The common eigenstates of the loop operators on $l_1$ can be labelled by anyon types: $|a\>,a=1,e,m,f$, with the property that the eigenvalue under action by $W^{a}_1$ is given by the braiding statistics of anyons: $W^{a}_1|b\>=e^{i\theta(a,b)}|b\>$. Here $\theta(a,b)$ is the braiding phase between anyon $a$ and $b$. The state $|a\>$ can be viewed as having an anyon $a$ threading through the center of the torus\footnote{In other words the state $|a\>$ can be viewed as prepared by a path integral in the solid torus, with an anyon string $a$ inserted in the center $S^1$.}, then the action of $W^b_1$ is performing a braiding between $a$ and $b$ in spacetime. The string-net basis is related to MES via a linear transformation. To derive the transformation, notice the effect of the operators $W^{e,m}_1$ on the string-net configuration: $W^e_1$ creates a loop along $l_1$ while $W^m_1$ measures the parity of loops along $l_2$. We can then deduce the action of $W^a_1$ on the string-net basis: 
 \begin{align}
    &W^e_1|0,0\>_{SN}=|1,0\>_{SN},~W^m_1|0,0\>_{SN}=|0,0\>,\\
    &W^e_1|1,0\>_{SN}=|0,0\>_{SN},~W^m_1|1,0\>_{SN}=|1,0\>_{SN},\\
    &W^e_1|0,1\>_{SN}=|1,1\>_{SN},~W^m_1|0,1\>_{SN}=-|0,1\>,\\
    &W^e_1|1,1\>_{SN}=|0,1\>_{SN},~W^m_1|1,1\>_{SN}=-|1,1\>_{SN},
\end{align}
from which we can form superpositions to obtain (unnormalized) eigenstates of loop operators $W^a_1$:
$|1\>=|0,0\>_{SN}+|1,0\>_{SN}$, $|m\>=|0,0\>_{SN}-|1,0\>_{SN}$, $|e\>=|1,1\>_{SN}+|0,1\>_{SN}$, $|f\>=|1,1\>_{SN}-|0,1\>_{SN}$. The inverse transformation is: $|0,0\>_{SN}=|1\>+|m\>$, $|1,0\>_{SN}=|1\>-|m\>$, $|0,1\>_{SN}=|e\>-|f\>$, $|1,1\>_{SN}=|e\>+|f\>$.

Now consider taking an MES $|a\>$ as our initial state and denote the post-measurement state as $P_m|a\>$ where $P_m$ is the projection operator for a measurement outcome $m$. The state $P_m|a\>$ remains topologically ordered in the unmeasured region $A,B$, since the stabilizers $A_v, B_p$ are not altered by the measurements for $v,p$ in $A\cup B$. A crucial observation is that the post-measurement state $P_m|a\>$ remains a common eigenstate of loop operators $W^a_1$ in region $A,B$. This is due to the fact that the loop operators $W^a_1$ are topological and can be freely deformed to be supported entirly in region $A$ or $B$ without changing their action on the states prior to measurement. Therefore since region $A,B$ are not affected by the measurements, the operators $W^a_1$ commute with the projector $P$, and $P_m|a\>$ remains an eigenstate of $W^a_1$ with the same eigenvalues as the initial state $|a\>$, that is, $W^b_1P_m|a\>=e^{i\theta(a,b)}P_m|a\>$

After measurement, the quantum state in $M$ becomes a trivial product state, therefore the $A,B$ to $M$ interfaces become gapped boundaries of the unmeasured region $A, B$. Region $A,B$ now host topological order on cylinder with gapped boundaries, whose ground space is finitely degenerated and can be characterized by the action of $W^a_1$s. Similar to the situation on a torus we can label the states in region $A,B$ by anyon types and they satisfy the relation $W^1_b|a\>=e^{i\theta(a,b)}|a\>$. One can again picture the state as having an anyon $b$ threading through the center of the cylinder. The fact that the post-measurement state $P_m|a\>$ remains an eigenstates of $W^b_1$ with eigenvalues $e^{i\theta(a,b)}$ fixes the states in region $A,B$ to be $|a\>_A,|a\>_B$. Therefore we conclude for an MES, 
\begin{align}
	\frac{P|a\>}{\sqrt{\<a|P|a\>}}= |a\>_A\otimes |a\>_B\otimes |\phi\>_M\label{eq: mesdec}
\end{align}
This shows for MES there is no MIE for any measurement basis. The decomposition of post-measurement state ~\eqref{eq: mesdec} allows us to directly calculate $\mie$ for a generic initial state by expanding the initial state in the MES basis. 

Let us reproduce the string-net states $\mie$ using the MES formalism. The state $|0,0\>_{SN}$ can be expanded in MES as $|0,0\>_{SN}=\frac{1}{\sqrt{2}}(|1\>+|m\>)$.  Apply the projection operator to the initial state and use the decomposition ~\eqref{eq: mesdec}, we conclude \footnote{Here we also used the fact that the norm square of the post measurement state $\<a|P|a\>$ is the same for $a=1,m$. This can be justified by noticing that $|a\>$s are related by the action of $W^a_2$s: $|a\>=W^a_2|1\>$, therefore $\<m|P|m\>=\<1|W^m_2PW^m_2|1\>$. Since $W^m_2$ is a string of $X$ operators that commutes with the measurements, we have $\<m|P|m\>=\<1|P|1\>$.  }
\begin{align}
	P|0,0\>_{SN}=\frac{1}{\sqrt{2}}(|1\>_A|1\>_B+|m\>_A|m\>_B)|\phi\>_M
\end{align}
which is a Bell-pair with entanglement entropy $\log 2$ regardless of measurement outcome. Therefore \mie~for $|0,0\>_{SN}$ is $\log 2$, in agreement with the result obtained earlier using the string-net picture of the ground state wavefunction.

\subsubsection{\mie~ in Abelian quantum double models}
The toric code analysis can be readily generalized to general Abelian topological orders. From the decomposition ~\eqref{eq: mesdec} we know the MES has zero MIE. Therefore a non-zero MIE can only be obtained if one starts with a non-MES state, such as the string-net states $|SN,\alpha,\beta\>$ of the toric code. 
We show that the string-net states of toric code can be generalized to generic abelian quantum double models $D(G)$, and their MIE contains information about the order of the gauge group $G$. 
\begin{figure}
	\centering
	\includegraphics[width=0.5\textwidth]{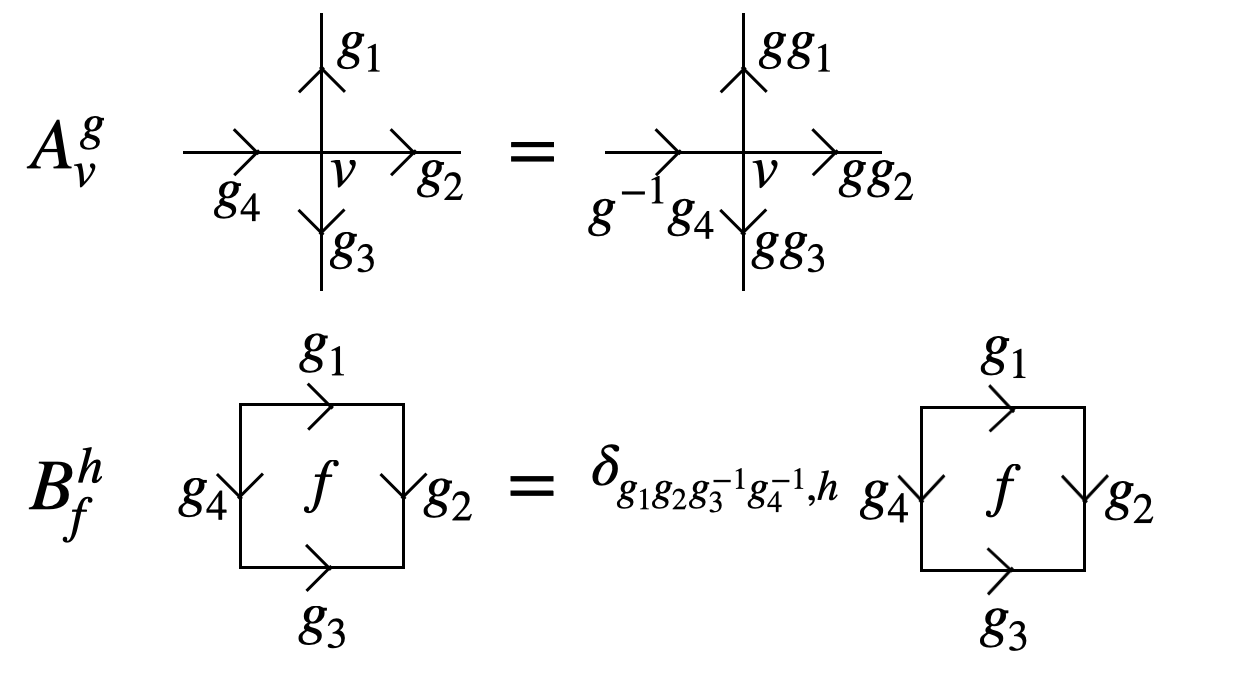}
	\caption{Definition of vertex and face operators in quantum double models.}
	\label{fig:TO_operator}
\end{figure}
The abelian quantum double model $D(G)$ with abelian $G$, can be defined on the square lattice as follows~\cite{Kitaev_2003,Bombin_2008}. There is a $|G|$-dimensional Hilbert space on every link of the lattice, with a natural basis $|g\>, g\in G$. The lattice is endowed with an orientation of edges and the Hamiltonian are built with two types of operators $A^g_v, B^h_p$, defined on vertices and faces of the lattice respectively. $A^g_v$ changes the states on the 4 edges adjacent to $v$ from $|h\>$ to $|gh\>$ or $|g^{-1}h\>$, according to whether the edge is pointing towards or away from the vertex $v$. $B^h_p$ forces the ``flux" through the plaquette $p$ to be $h$. These operators are summarized in Fig.~\ref{fig:TO_operator}. The Hamiltonian is then built from these two types of operators as follows:
 \begin{align}
	&H=-\sum_v A_v-\sum_f B_p,\label{eq:DGH}\\
	&A_v:=\frac{1}{|G|}\sum_g A^g_v,~ B_p:=B^e_p
 \end{align}
$A_v$ can be viewed of as the Gauss's law constraint and $B_f$ can be viewed as the zero-flux condition. The ground space of the quantum double can also be described by the picture of string-net condensation. However one needs to use a basis dual to $|g\>$ to express the ground state as a string-net liquid on the direct lattice.\footnote{Alternatively one can write the ground state of the quantum double as a string-net liquid in the $|g\>$-basis on the dual lattice. } For any character of the group $G$: $\chi\in \widehat{G}=\hom[G,U(1)]$, define a dual basis on an edge as $|\chi\>:=\frac{1}{\sqrt{|G|}}\sum_g \chi(g)|g\>$. This can be viewed as a group Fourier transformation of the basis $|g\>$, with inverse transformation $|g\>=\frac{1}{\sqrt{|G|}}\sum_{\chi\in \widehat{G}}\chi^*(g)|\chi\>$. Then acting on dual basis $|\chi_i\>$, where $i=1,2,3,4$ labels 4 edges adjacent to $v$, we have 
\begin{align}
	A_v\sim\frac{1}{|G|}\sum_g \chi_1^{s_1}(g)\chi^{s_2}_2(g)\chi^{s_3}_3(g)\chi^{s_4}_4(g),
\end{align}
where $s_i$ is the orientation of edge $i$ with respect to $v$. From properties of group Fourier transformation we have 
\begin{align}
	\frac{1}{|G|}\sum_g \chi_1^{s_1}(g)\chi^{s_2}_2(g)\chi^{s_3}_3(g)\chi^{s_4}_4(g)=\delta_{\prod_i \chi_i^{s_i},1}.
\end{align} 
Hence, the vertex term $A_v$ enforces the product of characters on edges adjacent to a vertex to be the trivial character. On the other hand the face term can be written in the $|g_i\>$ basis as $B_p\sim \delta_{g_1g_2g_3g_4,e}=\frac{1}{|G|}\sum_\chi \chi(g_1g_2g_3g_4)$. Therefore in the dual basis we have $B_p=\frac{1}{|G|}\sum_\chi B^\chi_p $, where $B^\chi_p$ changes the states on the the edges of the face $f$ from $|\chi_i\>$ to $|\chi \chi_i\>$. We see that in the dual basis the roles of $A_v$ and $B_p$ are interchanged. Then one can now view $A_v$ as enforcing the condition that in a ground state the characters flowing through any vertex is 1. $B_p$ terms then form a superposition of all possible such configurations. We can now say that the ground state of the Hamiltonian ~\eqref{eq:DGH} is a string-net liquid on the direct lattice, with string types labelled by characters of $G$. 

We are now in place to define a string-net basis for a general abelian quantum double model defined on a torus. In a given string-net configuration, the string type along a non-contractible direction $l_{1,2}$ is $\chi_{1,2}$. Then the ground states with definite $\chi_1,\chi_2$ are called string-net states and are denoted by $|\chi_1,\chi_2\>_{SN}$. We now make connection to the MES, defined for any topological order on a torus. The MES are labelled by anyon types. For abelian quantum double models, the anyon types are labelled by pairs $(g,\chi), g\in G, \chi\in \widehat{G}$, with self and mutual statistics given by $\theta(g,\chi)=\chi(g), \theta((g,\chi),(g',\chi'))=\chi(g')\chi'(g)$~\cite{Kitaev_2003,Bombin_2008}. $\chi$-type anyon is associated with violation of the vertex terms $A_v$ of the Hamiltonian, $g$-type anyon is associated with violation of face terms $B_p$. To find the transformation between MES $|g,\chi\>$ and string-net states $|\chi_1,\chi_2\>$, we analyze the effect of loop operators $W^1_{g,\chi}$ on the string-net states. The loop operator $W^1_{g,\chi}$ moves an anyon $(g,\chi)$ around $l_1$. It is an ribbon operator with support on a direct path and a dual path, both along $l_1$. It changes the states on the direct path from $|\chi_i\>$ to $|\chi\chi_i\>$, and the states on the dual path from $|g_i\>$ to $|gg_i\>$. Therefore it acts on the string-net states as $W^1_{g,\chi}|\chi_1,\chi_2\>_{SN}=\chi_2(g)|\chi\chi_1,\chi_2\>_{SN}$. To obtain MES, which are eigenstates of $W^1_{g,\chi}$, we simply perform Fourier transform to the first label of string-net states and define: $|g,\chi\>=\frac{1}{\sqrt{|G|}}\sum_{\chi_1}\chi^*_1(g)|\chi_1,\chi_2\>_{SN}$. One can verify the states $|g,\chi\>$ satisfies $W^1_{g',\chi'}|g,\chi\>=\chi_(g')\chi'(g)|g,\chi\>$. The phase $\chi(g')\chi'(g)$ is exactly the braiding between two anyons $(g,\chi)$ and $(g',\chi')$.Thus the states $|g,\chi\>$ are indeed the MES. The inverse transformation is given by the inverse Fourier transformation: $|\chi_1,\chi_2\>_{SN}=\frac{1}{\sqrt{|G|}}\sum_g \chi_1(g)|g,\chi_2\>$. 

Now using the decomposition of MES after measurement ~\eqref{eq: mesdec}, we have for string-net states 
\begin{align}
	P|\chi_1,\chi_2\>_{SN}=\frac{1}{\sqrt{|G|}}\sum_g \chi_1(g)|g,\chi_2\>_A\otimes |g,\chi_2\>\otimes |\phi\>_M
\end{align}
The entanglement entropy between $A$ and $B$ is therefore $\sum_{g\in G}\frac{1}{|G|}\log|G|=\log |G|$ regardless of measurement outcome. We conclude that for string-net states we have $\mie=\log G=\log D$. We used the fact that the quantum double $D(G)$ has total quantum dimension $D=|G|$.

\section{Discussion}
This exploration of measurement-induced entanglement (MIE) and information (MII) in ground-states of various systems reveals the presence of universal features depending on the topology or universality class for critical or gapless systems. This universal structure provides distinct, complementary information from that contained in the single-interval entanglement entropy of the state.
These quantities are relevant for assessing the computational power of the state for measurement-based quantum computing, and have operational significance for the classical and quantum complexity of simulating that state.
With the exception of gapped topological phases, our present understanding of MIE and MII comes entirely from numerical simulations of piecemeal examples. 
To obtain a systematic understanding of the structure of MIE and MII, analytic methods for computing these from field theory descriptions would be highly valuable, and are an important challenge for future work.
For example, while numerical simulations of the XX spin chain show that the MIE behaves like a four-point correlation function, although the scaling exponent does not appear to be simply related to known bulk- or boundary- scaling dimensions. This suggests that the MIE for $1+1d$ CFTs could, perhaps, be governed by new classes of scaling operators that have not been previously considered.
An analytic understanding would be especially valuable for studying higher-dimensional gapless systems, such as non-Fermi liquids, and gapless quantum spin liquids, where purely (classical) numerical methods are challenging to implement at large scale.

\vspace{6pt}\noindent{\it Acknowledgements -- } We thank Tim Hsieh and Ehud Altman for stimulating conversations. We also thank Yizhi You for pointing out typos in the manuscript. This work was supported by the US Department of Energy under awards DOE DE-SC0022102 (A.C.P.), the Air Force Office of Scientific Research under Grant No. FA9550-21-1-0123 (R.V.) and by a Sloan Research Fellowship (A.C.P.). We thank the Kavli Institute of Theoretical Physics (KITP) and the Aspen Center for Physics where part of this work were completed. KITP is supported in part by the National Science Foundation under Grant No. NSF PHY-1748958.

\bibliography{miebib}

\appendix

\section{Measurements in Gaussian fermion states}\label{ap:measurement_free_fermion}
A Gaussian fermion state can be entirely captured by its single particle correlation function $C_{ij}\equiv\langle c^\dagger_ic_j\rangle$~\cite{peschel2009reduced}. 
Here  we consider the effect of single orbital measurements on Gaussian fermion states, where the on-site projectors are given by 
\begin{align}
	P_1 = c_a^\dagger c_a, \quad
	P_0 = c_a c_a^\dagger=1-c_a^\dagger c_a,
\end{align}
with probabilities
\begin{align}
	p_1 = C_{aa},\quad
	p_0 = 1-C_{aa}.
\end{align}
where  $a$ is the measured orbital.

When $P_1$ is applied, the updated correlation matrix is given by
\begin{align}
	C'_{ij} = \frac{\langle c^\dagger_a c_ac^\dagger_i c_ic^\dagger_a c_a\rangle}{C_{aa}}=\left\{\begin{aligned}\ &1&\mathrm{if}\ i=j=a\\ \ &
		C_{ij}-\frac{C_{ia}C_{aj}}{C_{aa}}&\mathrm{if}\ i\neq a,\ j\neq a \\ \ &0&\mathrm{otherwise}\end{aligned}\right.
\end{align}
and when $P_0$ is applied, the updated correlation matrix is given by
\begin{align}
C'_{ij} = \frac{\langle c_a c^\dagger_ac^\dagger_i c_ic_a c^\dagger_a\rangle}{1-C_{aa}}=\left\{\begin{aligned}\ &0&\mathrm{if}\ i=j=a\\
		\ &C_{ij}+\frac{C_{ia}C_{aj}}{1-C_{aa}}&\mathrm{if}\ i\neq a,\ j\neq a\\ \ &0&\mathrm{otherwise}\end{aligned}\right.
\end{align}
where multi particle correlators can be decomposed to single particle correlators by Wick's theorem.
With the update rule, one can easily obtain the postmeasurement correlation functions satisfying Born probability.

\section{Correlation matrix for the 1$d$ free fermion chain}\label{ap:correlation_function}
In this section we derive the correlation matrix in ~\eqref{eq;correlationmatrix}. For the 1$d$ free fermion chain, the Hamiltonnian can be writte in the momentum space $H=-\sum_k2\cos k c^\dagger_kc_k$. For a ground state with filling factor $n_f$, we have
\begin{align}
	\langle c^\dagger_kc_k\rangle=\left\{\begin{aligned}
	&1 \mathrm{\quad if}\  |k|<n_f\pi\\&0 \mathrm{\quad 
		otherwise}
	\end{aligned}\right..
\end{align}
Then the correlation matrix becomes
\begin{align}\label{eq:cij}
	C_{ij}=\langle c^\dagger_ic_j\rangle=\frac{1}{L}\sum_k\langle c^\dagger_kc_k\rangle e^{ik(i-j)}=\frac{1}{L}\sum_{\vert k\vert<n_f\pi}e^{ik(i-j)}.
\end{align}
Note that when applying Jordan-Wigner transformation to a spin model with periodic boundary condition, the mapped 1$d$ free fermion chain has periodic (antiperiodic) boundary condition when the total number of fermion is odd (even). Thus, we consider the correlation matrix for odd and even number of fermion respectively.

When total number of fermion, i.e. $n_fL$, is odd, the allowed momentum is in the form of $2n\pi/L$, where $n$ is integer. Then \eqref{eq:cij} becomes
\begin{align}
	C_{ij}=&\frac{1}{L}+\frac{2}{L}\sum_{1\leq n\leq\frac{n_fL-1}{2}}\cos\frac{2\pi n(i-j)}{L}\nonumber\\
	=&\frac{1}{L}+\frac{2}{L}\frac{\sin\frac{\pi(n_fL-1)(i-j)}{2L}}{\sin\frac{\pi(i-j)}{L}}\cos\frac{\pi(n_fL+1)(i-j)}{2L}\nonumber\\
	=&\frac{\sin\pi n_f(i-j)}{L\sin \frac{\pi(i-j)}{L}},
\end{align}
where we have used the identity
\begin{align}\label{eq:id1}
	\sum_{n=1}^m\cos nx = \frac{\sin\frac{mx}{2}}{\sin\frac{x}{2}}\cos\frac{(m+1)x}{2}.
\end{align}

When total number of fermion, i.e. $n_fL$, is even, the allowed momentum is in the form of $(2n-1)\pi/L$. Then \eqref{eq:cij} becomes
	\begin{align}
		C_{ij}=&\frac{2}{L}\sum_{1\leq n\leq\frac{n_fL}{2}}\cos\frac{\pi(2n-1)(i-j)}{L}\nonumber\\
	 =&\frac{2}{L}\sum_{1\leq n\leq\frac{n_fL}{2}}\cos\frac{2\pi n(i-j)}{L}\cos\frac{\pi(i-j)}{L}\nonumber\\
	 &+\frac{2}{L}\sum_{1\leq n\leq\frac{n_fL}{2}}\sin\frac{2\pi n(i-j)}{L}\sin\frac{\pi(i-j)}{L}\nonumber\\
		=&\frac{\sin\pi n_f(i-j)}{L\sin \frac{\pi(i-j)}{L}},
	\end{align}
	where we have used \eqref{eq:id1} and
	\begin{align}
		\sum_{n=1}^m\sin nx = \frac{\sin\frac{mx}{2}}{\sin\frac{x}{2}}\sin\frac{(m+1)x}{2}.
	\end{align}
In conclusion, \eqref{eq;correlationmatrix} holds for both odd and even total number of fermion.

\begin{figure}
	\centering
	\includegraphics[width=0.5\textwidth]{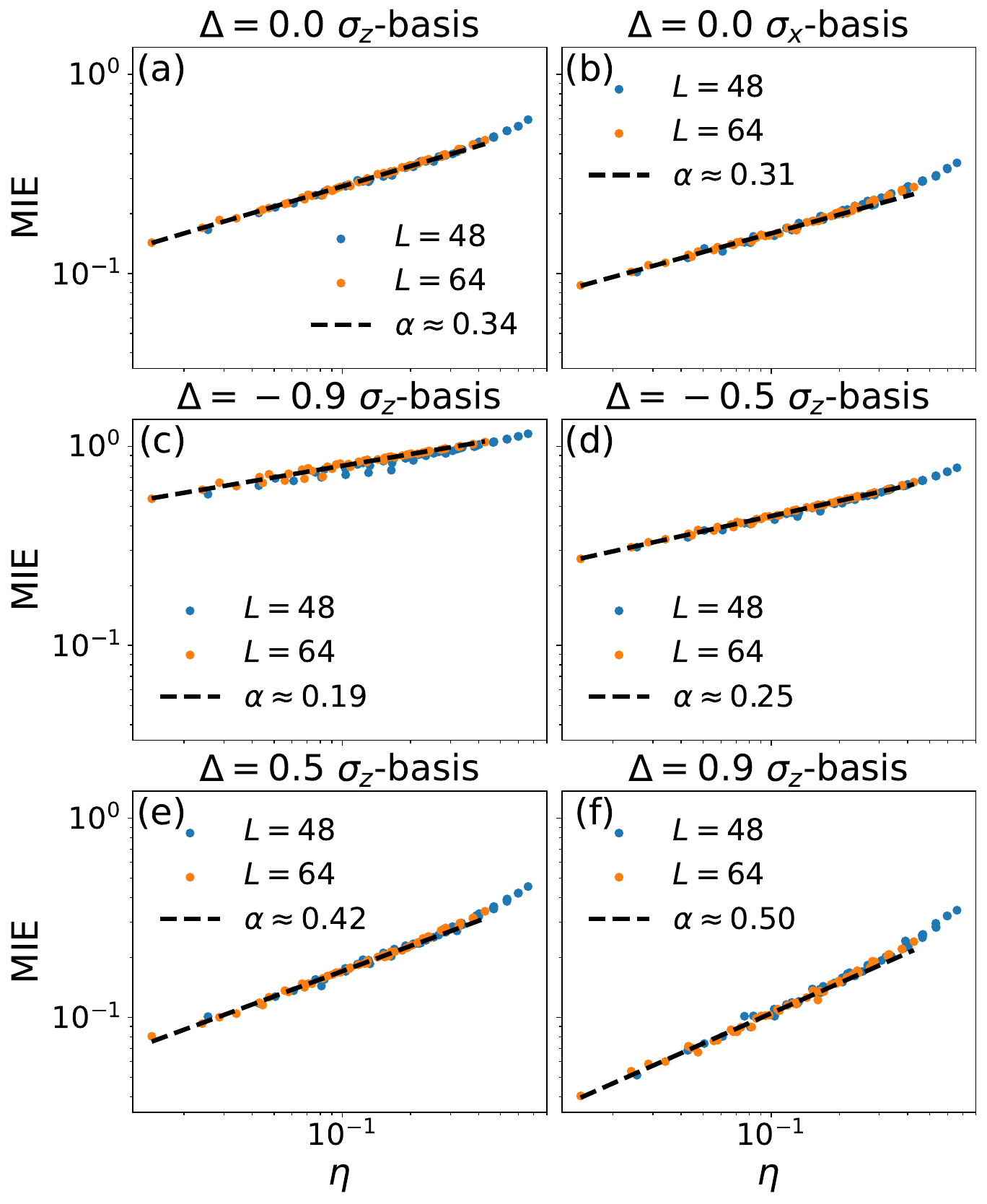}
	\caption{MIE calculated by DMRG for (a)/(b) the XX model with $\sigma_z$/$\sigma_x$-basis measurements  and for the XXZ model with $\sigma_z$-basis measurements and (c) $\Delta = -0.9$, (d) $\Delta = -0.5$, (e) $\Delta = 0.5$, and (f) $\Delta = 0.9$. Data are averged over from $3\cdot10^3$ to $6\cdot10^3$ measurement realizations.
	}
	\label{fig:MIE_XXZ}
\end{figure}

\section{DMRG calculation of MIE in the XX model with $\sigma_x$-basis measurements and the XXZ model}\label{ap:DMRG}
In this section we calculate the MIE in the XX and XXZ model with ground states obtained by the DMRG method implemented in the TeNPy package~\cite{tenpy}. The truncated bond dimensions are $256$ and $400$ for $L=48$ and $L=64$ respectively. As a benchmark, we first investigate the XX model with $\sigma_z$-basis measurements, which we have calculated via the free-fermion method up to $L=256$ in Sec.~\ref{ssec:gapless1d}. As shown in Fig.~\ref{fig:MIE_XXZ}(a), the fitted exponent $\alpha\approx0.34$ is slightly larger than the $\alpha\approx0.31$ (Fig.~\ref{fig:XX_MIE}) obtained by the free-fermion method. We interpret the discrepancy as a finite size effect. For the XX model with $\sigma_x$-basis measurements, the fitted exponent $\alpha\approx0.31$ (Fig.~\ref{fig:MIE_XXZ}(b)) slightly smaller than that with $\sigma_z$-basis measurements at the same system size, but the discrepancy is too small to tell whether they will drift to the same exponent at the large $L$ limit. 
For the XXZ model, Fig.~\ref{fig:MIE_XXZ}(c)-(f) show that the exponent $\alpha$ increases from $0.19$ to $0.25$ to $0.42$ to $0.5$, as the anisotropic interaction parameter $\Delta$ varies from $-0.9$ to $-0.5$ to $0.5$ to $0.9$. This trend is consistent with the analytically understood behavior in the mutual information where $\alpha_{\rm MI}=1-\frac{\arccos\Delta}{\pi}$~\cite{calabrese2009entanglement}.

\section{MIE upper-bounds strange correlators}\label{ap:strangecorrelatorMIE}
Mutual information between regions $A$ and $B$ upper bounds connected correlators between operators supported in these regions. 
In this section, we show that the MIE similarly upper bounds \emph{strange} correlators.
Consider a state $|\psi\>$ which is partitioned into three regions: $A$, $B$, and $C$:
\begin{align}
	\vert\psi\rangle = \sum_{a,b,c}\phi_{abc}\vert abc\rangle,
\end{align}
where $a$, $b$, and $c$ label the local degree of freedom in $A$, $B$, and $C$. We apply measurements on region $C$ with outcome $\vert m_c\rangle$, and then the resulting wavefunction is 
\begin{align}
	\vert\psi_c\rangle=\frac{\langle m_c\vert\psi\rangle}{\sqrt{p_{m_c}}}=\sum_{a,b}\frac{\phi_{abm_c}}{\sqrt{p_{m_c}}}\vert ab\rangle,
\end{align}
with probability $p_{m_c}=\sum_{a,b}\vert\phi_{abm_c}\vert^2$. The averaged MIE in $A$  is defined as
\begin{align}
	\overline{\mie}(A, B) = \sum_{m_c}p_{m_c}S_A\(\vert\psi_c\rangle\).
\end{align}
In the same setting, we can also define a strange correlator between a post-measurement product state $\vert m\rangle=\vert m_am_bm_c\rangle$  and the interested state $\vert\psi\rangle$ 
\begin{align}
	\mathrm{SC}\(O_A, O_B\)=\frac{\langle m\vert O_AO_B\vert\psi\rangle}{\langle m\vert \psi\rangle}
\end{align}
where $m_a$ and $m_b$ are meaurement outcomes in regions $A$ and $B$, which are not measured in the MIE setting, and $O_A$ and $O_B$ are charged local operators in regions $A$ and $B$. 

To show the connection between strange correlators and MIE, we first rewrite the MIE as the mutual information between $A$ and $B$ in the post-measurement state $\vert\psi_c\rangle$
\begin{align}\label{aeq:MIE_MI}
	\mie(A, B)=\frac{1}{2}\sum_{m_c}p_{m_c}I(A, B)\[\vert\psi_c\rangle\].
\end{align}
Then we can follow Ref.~\cite{wolf2008area} to express the mutual information as a relative entropy
\begin{align}
	I(A,B)\[\vert\psi_c\rangle\]=S\(\rho^c_{AB}\vert \rho^c_A\otimes\rho^c_B\),
\end{align}
where $\rho_{AB}^c=\vert\psi_c\rangle\langle\psi_c\vert$ and $\rho_{A/B}^c=\tr_{B/A}\(\vert\psi_c\rangle\langle\psi_c\vert\)$. Then using the norm bound~\cite{ohya1993quantum} $S(\rho\vert\sigma)\geq\frac{1}{2}\vert\vert\rho-\sigma\vert\vert^2_1$ and the trace inequality $\vert\vert X\vert\vert_1\geq\tr\(XY\)/\vert\vert Y\vert\vert_{\infty}$, where $\vert\vert\cdot\vert\vert_p$ is the Schatten norm, we can obtain
\begin{align}
	I(A,B)\[\vert\psi_c\rangle\]\geq\frac{1}{2}\frac{\(\tr\(\rho_{AB}^cY\)-\tr\(\rho_A^c\otimes\rho_B^cY\)\)^2}{\vert\vert Y\vert\vert_{\infty}^2}.
\end{align}
Let $Y=\vert m_am_b\rangle\langle m_am_b\vert O_AO_B$, then the right hand side of the inequality becomes
\begin{align}~\label{aeq:SC}
	\frac{p_{m_am_bm_c}^2}{2\vert\vert Y\vert\vert_{\infty}^2p_{m_c}^2}\[\mathrm{SC}(O_A, O_B)\]^2
\end{align}
where we have used the fact that each changed local operator has zero expectation value.

Combining ~\eqref{aeq:MIE_MI}--~\eqref{aeq:SC}, we can obtain
\begin{align}
	\mie(A, B)\geq c_0\sum_{m_c}\frac{p^2_{m_am_bm_c}}{p_{m_c}}\[\mathrm{SC}(O_A, O_B)\]^2
\end{align}
where $c_0$ is some non-universal constant depending on the choice of $\vert m_{a/b}\rangle$ and $O_{A/B}$. Therefore, MIE is lower-bounded by the average of square of strange correlators weighted by some nonuniversal joint measurement probability.
\end{document}